\documentclass[11pt,a4paper]{article}
\pdfoutput=1

\usepackage{hyperref}
\usepackage{jheppub}
\usepackage{subfigure}
\usepackage{slashed}

\newcommand{\ds}{\displaystyle×}

\newcommand{\ra}{\rightarrow}
\newcommand{\tr}{\tilde{t}_R}
\newcommand{\eref}[1]{Eq.(\ref{#1})}
\newcommand{\fref}[1]{FIG. \ref{#1}}

\title{Top Quark Forward-Backward Asymmetry and Anomalous Right-Handed Charge Currents}

\author[a]{John N. Ng}
\author[a,b]{and Peter T. Winslow}

\affiliation[a]{Theory Group, TRIUMF, 4004 Wesbrook Mall, Vancouver, BC, V6T 2A3, Canada}
\affiliation[b]{Department of Physics and Astronomy, University of British Columbia, Vancouver, BC, V6T 1Z1, Canada}

\emailAdd{misery@triumf.ca}
\emailAdd{pwinslow@triumf.ca}

\abstract{Recent measurements of the top quark forward-backward asymmetry at the Tevatron could hint at new physics with an unexpected flavor structure. The significance of such an abnormal flavor structure in alleviating the tension between the inclusive and exclusive determinations of $|V_{ub}|$ via right-handed charge currents is studied. In particular, we elaborate on how the associated new flavor changing couplings naturally allow for the generation of anomalous loop-induced right-handed charge currents which can simultaneously remove the tension in the determinations of $|V_{ub}|$ and escape the tight indirect bounds from $B \ra X_s \gamma$.}

\keywords{Beyond Standard Model, B-Physics, Heavy Quark Physics, Phenomenological Models}

\begin{document}
\maketitle
\flushbottom

\section{Introduction}

The top quark forward backward asymmetry (FBA) has been measured by both D$\slashed{0}$ and CDF at the Fermilab Tevatron collider \cite{:2007qb,Aaltonen:2008hc,Aaltonen:2011kc,Abazov:2011rq}. In their most recent measurement the CDF collaboration found a dramatic dependence of the size of the asymmetry on the $t \overline{t}$ invariant mass $M_{t \overline{t}}$. In particular, in the high invariant mass region, $M_{t \overline{t}} > 450 \text{ GeV}$, the parton level asymmetry in the $t \overline{t}$ rest frame was found to deviate from the next to leading order (NLO) standard model (SM) prediction by 3.4$\sigma$. However, in the low invariant mass region, $M_{t \overline{t}} < 450 \text{ GeV}$, the deviation reduced to only $\sim 1 \sigma$. Although the D$\slashed{0}$ collaboration does confirm the possibility of an interesting deviation from the SM in the $t \overline{t}$ FBA they find no statistically significant enhancement of the asymmetry in the high invariant mass region. 

These results have inspired many theoretical models to explain the anomaly, e.g.,  \cite{Ferrario:2009bz,Frampton:2009rk,Chivukula:2010fk,Gresham:2011dg,Shelton:2011hq,Djouadi:2009nb,Alvarez:2010js,Chen:2010hm,Zerwekh:2011wf,Barreto:2011au,Jung:2011ue,Jung:2009jz,Cheung:2009ch,Shu:2009xf,
Dorsner:2009mq,Arhrib:2009hu,Jung:2009pi,Barger:2010mw,Gupta:2010wt,Xiao:2010hm,Jung:2010yn,Cao:2011ew,Berger:2011ua,Cao:2010zb,Barger:2011ih,Grinstein:2011yv,Patel:2011eh,Bhattacherjee:2011nr,Ko:2011vd,Ko:2011di}. One particular class of models which has received a large amount of attention involves new particle content exchanged in the $t$-channel of the $t \overline{t}$ pair production process \cite{Jung:2009jz,Cheung:2009ch,Shu:2009xf,
Dorsner:2009mq,Arhrib:2009hu,Barger:2010mw,Gupta:2010wt,Xiao:2010hm,Cao:2011ew,Berger:2011ua,Cao:2010zb,Barger:2011ih,Grinstein:2011yv,Patel:2011eh,Bhattacherjee:2011nr}. The main signature of these models is destructive interference with the SM QCD amplitudes stemming from large inter-generational couplings which connect the $u$ and $t$ quarks via some new mediator (scalar or vector). In order to produce the desired interference effects the coupling must be chiral and, in the case of a vector mediator, is generally chosen to be right-handed in order to avoid strong constraints from electroweak precision data and flavor physics. Considering the strong constraints on flavor changing neutral currents (FCNC) in the down quark sector this mediator should either only couple to up-type quarks or have flavor diagonal couplings in the down quark sector. Even in the latter case, care must be taken as flavor diagonal couplings to $b \overline{b}$ could potentially lead to undesirable deviations in precision electroweak observables. 

It has been known for some time now that there exists tension between the inclusive and exclusive determinations of $|V_{ub}|$ \cite{Charles:2004jd,Nakamura:2010,Bona:2009cj,Stone:2011vd}. Experimental determinations of $|V_{ub}|$ are based on the semi-leptonic decay rates of the B meson. The inclusive value of $|V_{ub}|$, $|V_{ub}|_{incl}$, is extracted from measurements in which only the final state lepton is detected, $B \ra X_u \ell \nu$. This measurement is made difficult by the large charmed background owing to the fact that $|V_{ub}|<<|V_{cb}|$. Due to this, a proper extraction of $|V_{ub}|$ requires one to either look directly for exclusive final states, in which case the extraction yields the exclusive value $|V_{ub}|_{excl}$, or suppress the charmed background kinematically. The inclusive and exclusive results do not agree, with $|V_{ub}|_{incl} = (4.25 \pm 0.15_{exp} \pm 0.20_{th}) \times 10^{-3}$ \cite{Kowalewski} and $|V_{ub}|_{excl} = (3.25 \pm 0.12_{exp} \pm 0.28_{th}) \times 10^{-3}$ \cite{Urquijo} which differ by $\sim 25$\%. In the exclusive case, one needs knowledge of the relevant hadronic matrix elements while, in the inclusive case, there can be difficulties in terms of understanding the effect of the necessary kinematic cuts on the perturbative expansion. In both cases there can be a large amount of uncertainty involved. The branching ratio for $B \ra \tau \nu$ is also a useful observable in the study of the different determinations of $|V_{ub}|$. In the SM the branching ratio is given by 

\begin{equation}
\text{Br} ( B \ra \tau \nu ) = \frac{G_F^2 m_B m_\tau^2}{8 \pi} \left( 1 - \frac{m_\tau^2}{m_B^2} \right)^2 f_B^2 |V_{ub}|^2 \tau_B
\label{eq:1}
\end{equation}
where the Fermi constant $G_F$, the $B$ meson and $\tau$ lepton masses, and the $\tau$ lifetime have all been precisely measured \cite{Nakamura:2010} leaving only the decay constant $f_B$ and $|V_{ub}|$. If the lattice determination of the decay constant, $f_B = 208 \pm 8 \text{ MeV}$  \cite{Gamiz:2009ku}, is taken as reliable then the branching ratio is essentially a function of $|V_{ub}|$ only. The branching ratio is within 1$\sigma$ of the experimental value for $|V_{ub}|_{incl}$ while it deviates by $\sim$3$\sigma$ for $|V_{ub}|_{excl}$ \cite{Lunghi:2010gv}. Although this deviation can be caused by some combination of the above mentioned uncertainties we consider the case in which it is caused by the effects of so-far unknown new physics (NP). Considering the proximity of the prediction of $\text{Br}(B \ra \tau \nu)$ to the experimental value using $|V_{ub}|_{incl}$, this seems to imply that the NP should enter distinctively into the exclusive decays.

Recently it has been realized that anomalous right-handed charge currents (RHCC) in the quark sector can alleviate the tension between the different determinations of $|V_{ub}|$ while simultaneously improving agreement between the prediction and experimental value of $\text{Br}(B \ra \tau \nu)$ \cite{Chen:2008se,Crivellin:2009sd,Buras:2010pz}. Most recently, the authors of \cite{Buras:2010pz} considered an effective RHCC and corresponding mixing matrix, $V^R$, generated by an enlargement of the gauge symmetry. The interference effects of the right-handed mixing matrix $V^R$ with the standard left-handed mixing matrix, denoted as $V^L$, were suppressed in the inclusive decays $B \ra X_u \ell \nu$ but not in the exclusive decay $B \ra \pi \ell \nu$, implying that the NP dominantly affects the exclusive decay channels. A global fit to the elements $V^L_{ub}$ and $V^R_{ub}$ using the experimental constraints from the semi-leptonic inclusive decay $B \ra X_u \ell \nu$, the semi-leptonic exclusive decay $B \ra \pi \ell \nu$, and the purely leptonic decay $B \ra \tau \nu$ determined the best fit values $|V^L_{ub}| = (4.1 \pm 0.2) \times 10^{-3}$ and $\textrm{Re} ( V^R_{ub}/V^L_{ub} ) = -0.19 \pm 0.07$. For comparison, the most recent inclusive measurement of $|V_{ub}|$ is $|V_{ub}|_{incl} = (4.25 \pm 0.15_{exp} \pm 0.20_{th}) \times 10^{-3}$ \cite{Kowalewski}. 

In the standard model the tree level $W$ boson couplings to fermions are strictly left-handed by construction. However, small effective RHCC are generated though quantum loop effects. The relative strength of such currents in the SM is $\mathcal{O}  \left( \ds \frac{\alpha_g}{4\pi} \frac{m_f m_{f'} }{M_W^2} \right)$ where $m_{f}$, $m_{f'}$, and $M_W$ are the masses of the fermions and the $W$ boson respectively and $\alpha_g$ is the relevant coupling. Even for RHCC involving the top quark the largest correction is due to QCD and the effect is expected to be less than 1\%. Many NP models can induce sizeable tree level right-handed currents by enlarging the gauge symmetry \cite{Pati:1974vw,Mohapatra:1974gc,Mohapatra:1974hk,Senjanovic:1975rk,Senjanovic:1978ev,Agashe:2003zs,Zhang:2007da,Maiezza:2010ic}. However, severe constraints from the active neutrino sector generally force the scale of these models to inaccessibly high energies. An alternative route is to generate RHCC via loop effects involving new beyond SM particle content. This method has the added benefit that, as long as the active neutrinos are left-handed and the right-handed sterile neutrinos are sufficiently heavy, no leptonic right-handed currents can be generated; thereby relegating all NP effects to the quark sector and avoiding the stringent constraints associated with the leptonic sector. 

In the current work we explore the role of the abnormal flavor structure hinted at by the top quark FBA in enhancing the strength of anomalous loop-induced RHCC contributing to $B \ra X_u \ell \nu$ decays and therefore ameliorating the tension in the different determinations of $|V_{ub}|$. The same flavor couplings which enhance RHCC in B decays also have the potential to generate large loop-induced RHCC in top decays which are subject to strong indirect constraints from rare $B \ra X_s \gamma$ decays. Despite this, we will show that RHCC generated by the virtual exchange of a new scalar (vector) in top decays are CKM (as well as chiral) suppressed while in B decays the RHCC are CKM (and chiral) enhanced thereby allowing us to choose large couplings to alleviate the $|V_{ub}|$ tension and remain unconstrained by the strong indirect $B \ra X_s \gamma$ constraints.

The layout of the paper is as follows: in section II we describe the effective formalism used, briefly review the relevant indirect constraints from $B \ra X_s \gamma$, and discuss the necessary bounds on the right-handed mixing matrix to alleviate the $|V_{ub}|$ tension. In section III we examine the case of a leptophobic $Z'$ which mediates the flavor changing coupling, while in section IV we examine the possibility that the flavor couplings are associated with a scalar mediator. In section V we discuss how flavor off-diagonal elements of squark mass matrices in the context of the MSSM and a recently proposed supersymmetric explanation of the top quark FBA can also generate the RHCC and in section VI we conclude.

\section{Formalism}

 In the language of effective field theory, the effects of any anomalous loop-induced chiral charge currents are felt strictly through the presence of gauge invariant dimension six operators of the type

\begin{eqnarray}
c^R_{ij} \mathcal{O}^{R}_{ij} &=& \ds \frac{c^R_{ij}}{\Lambda^2} \overline{u}_{R i} \gamma^\mu d_{R j}  \tilde{H}^{\alpha \dagger} \left( i D_\mu H \right)_\alpha + h.c. \notag \\
c^L_{ij} \mathcal{O}^{L}_{ij} &=& \ds \frac{c^L_{1 ij}}{\Lambda^2} \overline{Q}_{L i}^{\alpha} H^{\beta \dagger} \left( i \slashed{D} H \right)_\alpha Q_{j \beta L} \notag \\
&+& \ds \frac{c^L_{2 ij}}{\Lambda^2} \overline{Q}^\alpha_{i L} \gamma^\mu \left( \tau^a \right)_\alpha^{\; \; \beta} Q_{j \beta L} H^{\gamma \dagger} \left( \tau^a \right)_\gamma^{\; \; \delta} \left( i D_\mu H \right)_\delta  + h.c. 
\label{eq:2}
\end{eqnarray}
where $u_R$, $d_R$, and $Q_L$ are the $SU(2)_L$ singlet right-handed up and down type quarks and the left-handed $SU(2)_L$ quark doublets respectively. Flavor indices are denoted as $i,j$ while the fundamental and adjoint $SU(2)_L$ indices are denoted as $\alpha, \beta, \gamma, \delta$ and $a$ respectively. The SM Higgs doublet is $H$ while $\tilde{H}=i\sigma_2 H^*$ and the scale of the NP that gives rise to these new operators is denoted by $\Lambda$. After spontaneous symmetry breaking the diagonalization of the quark mass terms is achieved by a set of bi-unitary rotations given by

\begin{eqnarray}
u'_{i L} \ra A^u_{ij} u_{j L} & \hspace{.4in} & d'_{i L} \ra A^d_{ij} d_{j L} \notag \\
u'_{i R} \ra B^u_{ij} u_{j R} & \hspace{.4in} & d'_{i R} \ra B^d_{ij} d_{j R}  .
\label{eq:3}
\end{eqnarray}

The effective operators in \eref{eq:2} then alter the charged current sector of the SM lagrangian such that 

\begin{eqnarray}
\mathcal{L}_{W^\pm} = e_W   \overline{u}_i \gamma^\mu \left(  V^L_{ij}  P_L + V^R_{ij} P_R \right) d_j W_\mu^+ + h.c. 
\label{eq:4}
\end{eqnarray}
where $e_W = g/\sqrt{2}$, $g$ is the $SU(2)_L$ gauge coupling, and $P_{L,R} = 1/2(1 \mp \gamma^5)$. The matrix $V^L_{ij}$ has contributions from the tree level SM CKM matrix and also from the operator $\mathcal{O}^L_{ij}$ whereas $V^R_{ij}$ consists solely of the $\mathcal{O}^R_{ij}$ contribution

\begin{eqnarray}
V^L_{ij} &=& V^{CKM}_{ij} + \frac{v^2}{2 \Lambda^2} \left( A^{u \dagger} c^L_1 A^d \right)_{ij} + \frac{v^2}{\Lambda^2} \left( A^{u \dagger} c^L_2 A^d \right)_{ij} \notag \\
V^R_{ij} &=& \frac{v^2}{2 \Lambda^2} \left( B^{u \dagger} c^R B^d \right)_{ij} 
\label{eq:5}
\end{eqnarray}
and $V^{CKM}_{ij} = \left( A^{u \dagger} A^d \right)_{ij}$. The SM charged current is reproduced by setting $c^L_{1 ij} = c^L_{2 ij} = c^R_{ij} = 0$. 

The effects of RHCC in top decay have been investigated in terms of their effects on observables which are sensitive to the $t \overline{b} W^-$ vertex structure. In particular, the branching ratio of the rare $B \ra X_s \gamma$ decays strongly, although indirectly, constrain any alterations to this structure. The constraints on $V^R_{tb}$ are particularly tight due to an enhancement by a factor $m_t/m_b$ \cite{Fujikawa:1993zu}. A relatively recent study has determined that these constraints imply the 95\% C.L. upper and lower bounds $-0.0007 \leq V^R_{tb} \leq 0.0025$ \cite{Grzadkowski:2008mf,Drobnak:2011aa} as long as one assumes no other anomalous couplings. If one drops these assumptions the possibility of cancellations which could loosen the constraints arise. We will assume that these bounds are robust and that their implications should be taken seriously.

In \cite{Buras:2010pz}, bounds on both the left and right-handed mixing matrices were determined by comparison with SM processes in $s \ra u$, $b \ra c$, and $b \ra u$ transitions. The results of the fits to the left-handed mixing matrix are

\begin{eqnarray}
|V^L_{us}| = 0.2248 \pm 0.0009 \hspace{.25in} |V^L_{cb}| = (40.7 \pm 0.6) \times 10^{-3} \hspace{.25in} |V^L_{ub}| = (4.1 \pm 0.2) \times 10^{-3}  . \notag \\
\label{eq:6}
\end{eqnarray}

We assume a Wolfenstein parametrization of the left-handed mixing matrix such that the first two equations in Eq. \ref{eq:6} uniquely determine the parameters $\lambda$ and $A$. In order to determine the rest of the Wolfenstein parameters we note that a large CP-violating phase in $B_s$ mixing, as hinted at by recent Tevatron experiments, can be accommodated partly by assuming that $|V^R_{td}| \sim 0$ \cite{Buras:2010pz}. In this limit, we can attribute the measured value $|V^{exp}_{td}| = (8.4 \pm 0.6) \times 10^{-3}$ \cite{Nakamura:2010} completely to the left-handed mixing matrix which, in combination with the final equation in Eq. \ref{eq:6}, determines the full set of Wolfenstein parameters. Using the set of parameters obtained in this way, we can then determine $\text{Re} ( V^L_{ub} ) = A \lambda^3 \rho = (1.6 \pm 0.5) \times 10^{-3}$. This leads to a range of allowed values for the real part of the corresponding element of the right-handed mixing matrix 

\begin{equation}
\text{Re} ( V^R_{ub} ) = (-3.1 \pm 1.5) \times 10^{-4} .
\label{eq:7}
\end{equation}

In what follows, we will interpret these bounds as the necessary bounds on the strength of the loop-induced RHCC to alleviate the $|V_{ub}|$ tension.

\section{A Leptophobic $Z'$}

Our spin 1 benchmark model is that of a new neutral $Z'$ boson. This is a well studied extension of the SM and can arise naturally in beyond SM scenarios which involve the breaking of new gauge symmetries. A leptophobic version of the $Z'$ model has been studied recently in the context of the $t \overline{t}$ asymmetry \cite{Jung:2009jz,Cao:2011ew,Berger:2011ua,Cao:2010zb,AguilarSaavedra:2011zy,AguilarSaavedra:2011ug,AguilarSaavedra:2011hz}. In this particular version the $Z'$ has a flavor changing, chiral coupling to $u_R$ and $t_R$. Any coupling to $u_L$ and $t_L$ must necessarily involve an interaction between the first and third generation quark doublets which, in the mass basis, leads to a coupling of the form $f_L V^*_{ud} V_{tb} \overline{d}_L \gamma^\mu b_L Z'_\mu$. Constraints on the mass difference between $B^0_d$ and $\overline{B}^0_d$ restrict the size of the left handed coupling such that $f_L < 3.5 \times 10^{-4} \left( M_{Z'} / 100 \text{ GeV} \right)$ \cite{Cao:2010zb} and it is therefore usually neglected completely. If the $Z'$ only interacts with $u_R$ and $t_R$ then its only available decay mode is $Z' \ra \bar{t}_R u_R, \bar{u}_R t_R$ which will lead to too many same sign top events at the Tevatron through the production mechanisms $u \bar{u}  \ra Z' Z' \ra tt \bar{u} \bar{u}$ and $g u \ra t Z' \ra t t \bar{u}$. In order to avoid this, a small flavor diagonal coupling to up-type quarks, $\epsilon_U \bar{u}_{i,R} \gamma^\mu u_{i, R} Z'_\mu$, is usually introduced \cite{Jung:2009jz,Cao:2011ew} to open up other more favorable decay modes. As long as the $Z'$ is lighter than the top quark the new flavor diagonal decay mode to $u \bar{u}$ will be the preferred one. However, $\epsilon_U$ cannot be too large either as it is subject to dijet constraints at the Tevatron and can, in conjunction with the large $t \bar{u}$ coupling, enhance the rate for the normally GIM and loop suppressed rare top decay $t \ra g u$. Additionally, if the $Z'$ is much lighter than the top quark ($M_{Z'} \lesssim 120 \text{ GeV}$) then this could lead to a large branching ratio for the decay $t \ra Z' u$ which would manifest itself as a large difference in the measurement of the total $t \bar{t}$ production cross section as measured in the lepton + jets and dilepton channels. A greater proximity of the $Z'$ mass to the top quark mass will suppress the $t \ra Z' u$ branching ratio and prevent these large discrepancies. 

Aside from the Tevatron, the same coupling which generates the $t \bar{t}$ asymmetry also necessarily leads to ample production of same-sign top quarks at the LHC. The limits on the flavor changing coupling and the $Z'$ mass from requiring consistency with a large FBA and the measured $t \bar{t}$ production cross section at the Tevatron were investigated in \cite{Berger:2011ua,AguilarSaavedra:2011zy,AguilarSaavedra:2011ug} for a heavy $Z'$ ($M_{Z'} \ge 200 \text{ GeV}$). Following this, the CMS collaboration searched for same-sign $tt$ production and concluded that the heavy $Z'$ exchange explanation of the FBA is disfavored at greater than 2$\sigma$ \cite{arXiv:1106.2142}. We therefore focus our attention here on the still phenomenologically viable light $Z'$ explanation of the Tevatron FBA ($M_{Z'} \lesssim 200 \text{ GeV}$).

Although a flavor diagonal coupling to lighter up type quarks was previously employed to avoid too many same sign top events at the Tevatron, we see no reason why this flavor diagonal coupling need be in the up-type quark sector. In particular, since we are interested in the effect of such a flavor changing $Z'$ on $b \ra u$ transitions in $B$ meson decays we instead opt for a flavor diagonal coupling to $b \bar{b}$. In addition to providing an alternative decay mode for the $Z'$ in order to avoid too many same sign top quark pairs at the Tevatron this coupling also avoids dijet constraints by relying on parton luminosity suppression and does not lead to any enhancements of the rare top decay $t \ra g u$. It is, however, constrained by observables associated with $Z \ra b \bar{b}$ and rare $B \ra X_s \gamma$ decays. All of these constraints are discussed below. Aside from the constraints, the introduction of the new flavor diagonal coupling also leads to an interesting signature of single top quark production with an associated $b \bar{b}$ resonance which may be searched for at the LHC or in the existing data sets at the Tevatron. Specifically, we consider the following phenomenologically motivated lagrangian

\begin{equation}
\mathcal{L} = g_{utZ'} \overline{u} \gamma^\mu P_R t Z'^{\mu} + g_{\overline{b} b Z'} \overline{b} \gamma^\mu P_R b Z'^\mu + h.c.
\label{eq:8}
\end{equation}

The right-handed coupling to $b \bar{b}$ was chosen to loosen the constraints from the observable $R_b$ which will be discussed shortly. Note that these types of couplings need not be generated by charging standard model fields under a new $U'(1)$. Instead the $Z'$ can couple to the quark fields via higher dimensional effective operators \cite{Fox:2011qd}. 

In principle, the new $Z'$ is free to undergo kinetic mixing with the hypercharge gauge boson which, after spontaneous symmetry breaking, will induce $Z$-$Z'$ mixing. Adherence to constraints from electroweak precision data (EWPD) and the need to avoid the generation of any large effective leptonic couplings to the $Z'$ demand that such mixing be highly suppressed ($ \lesssim 10^{-3}$ \cite{Chang:2006fp}). We therefore treat this mixing as negligible in what follows. This corresponds to choosing the tree level value of the mixing such that the full amplitude (tree level plus 1-loop) falls below the necessary bound.

\begin{figure}[t!]
\centerline{\includegraphics[width=3in, height=2.75in, angle=0]{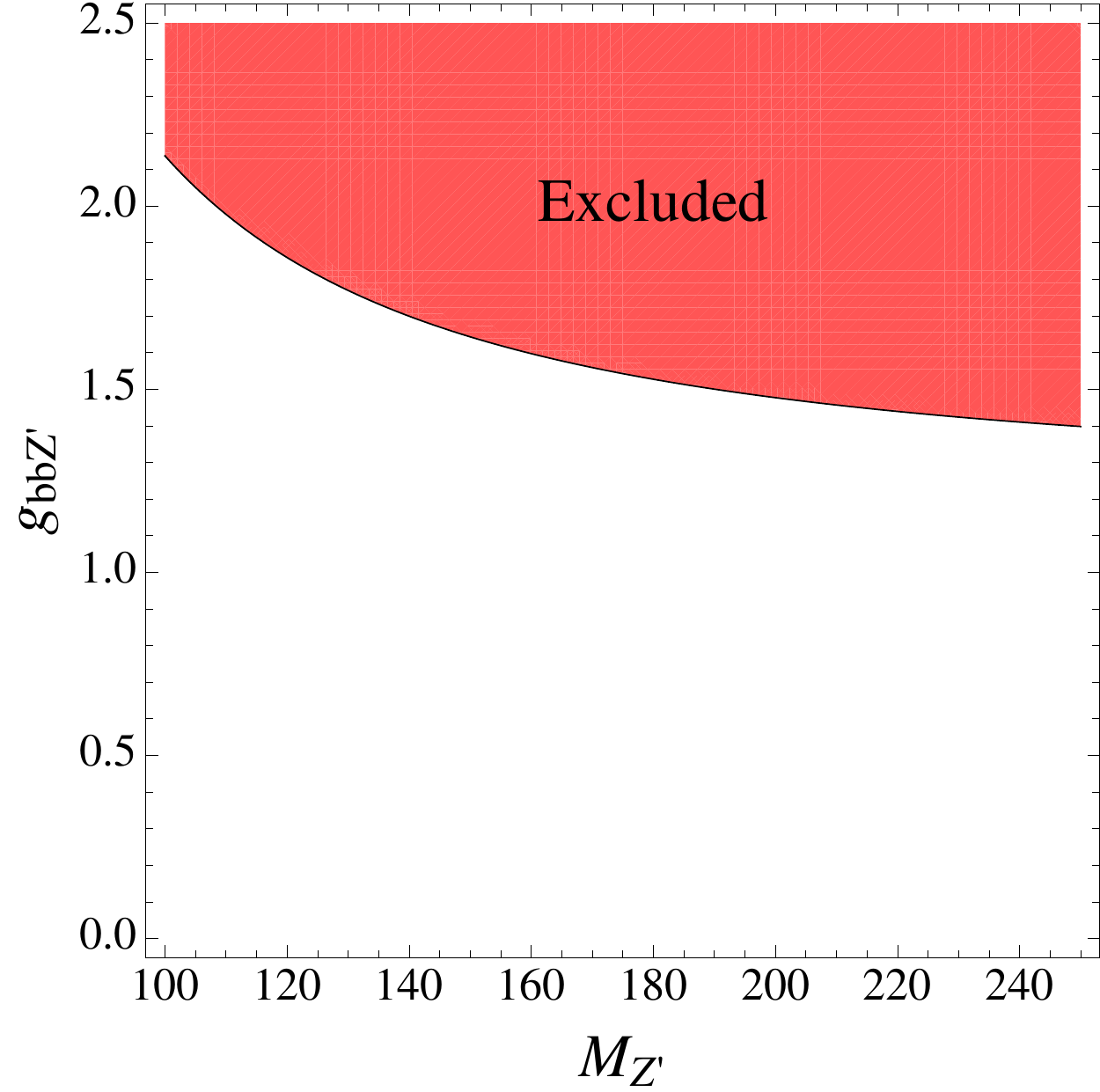}}
\caption{The red region is excluded by constraints from the observable $R_b$.}
\label{FIG1a}
\end{figure}

An important constraint on the mass of the $Z'$ and the coupling $g_{b b Z'}$ is the observable $R_b$, associated with the decay $Z \ra b \overline{b}$. The introduction of the flavor diagonal coupling $g_{bbZ'}$ modifies the $Z b \overline{b}$ vertex at the 1-loop level, shifting the SM tree level couplings such that $g_{b_{L,R}} \ra g_{b_{L,R}}^{SM} + \delta g_{b_{L,R}}$. These shifts can be detected through their effects on the observable $R_b$, defined as 

\begin{equation}
R_b = \frac{\Gamma ( Z \ra b \overline{b} )}{\Gamma ( Z \ra \text{Hadrons} )} = \frac{  (g_{b_L}^{SM})^2 + (g_{b_R}^{SM})^2 }{ \sum_{i=u,d,c,s,b} \bigg( (g_{i_L}^{SM})^2 + (g_{i_R}^{SM})^2 \bigg) } .
\label{eq:9}
\end{equation}

It is standard to normalize the $Z \ra b \overline{b}$ width with the total hadronic width as this leads to the cancellation of many QCD and electroweak corrections, thereby magnifying the sensitivity to the NP. The corresponding shift in $R_b$ due to the shift in the tree level SM couplings is parametrized as 

\begin{eqnarray}
\delta R_b = R_b - R_b^{SM} = 2 R_b^{SM} (1- R_b^{SM} ) \frac{( g_{b_L}^{SM} \delta g_{b_L} + g_{b_R}^{SM} \delta g_{b_R} ) }{ \bigg( (g_{b_L}^{SM})^2+ (g_{b_R}^{SM})^2 \bigg)} 
\label{eq:10}
\end{eqnarray}
to first order in $\delta g_{b_{L,R}}$. Assuming the ratio $m_b^2/M_{Z'}^2$ is negligible, only the right-handed $Z b \overline{b}$ coupling is shifted due to the polarization of the $Z' b \overline{b}$ vertex. At the 1-loop level the overall shift in the coupling is comprised of contributions from $Z$-$Z'$ mixing through a $b$ quark loop, mass renormalization of the external $b$ quark lines, and the vertex correction. We neglect the $Z$-$Z'$ mixing and assume that the full effect of the mass renormalization diagrams is to shift the mass of the $b$ quark to its physical value. The calculation of the vertex correction is performed in the unitary gauge. The result is finite 

\begin{eqnarray}
\delta g_{b_R}  = - g_{b_R}^{SM} \frac{ \alpha_{b b Z'} }{ 2 \pi } \left[ \frac{5}{2} + \frac{1}{x_Z} - \bigg( \frac{3}{2} + \frac{1}{x_Z} \bigg) \ln x_Z + \bigg( 1+ \frac{2}{x_Z} + \frac{1}{x_Z^2} \bigg) \bigg( \ln x_Z \ln ( 1+ x_Z) + \text{Li}_2 (- x_Z) \bigg) \right] \notag \\
\label{eq:11}
\end{eqnarray}
where $x_Z = M_Z^2/M_{Z'}^2$ and $\alpha_{b \overline{b} Z'} = g_{b \overline{b} Z'}^2/ 4 \pi$. Using the SM calculated values $g_{b_{L}}^{SM} = -0.4208$, $g_{b_{R}}^{SM} = 0.0774$, $R_b = 0.21578 \pm 0.00010$, and the measured value $R_b^{exp} = 0.21629 \pm 0.00066$ \cite{Nakamura:2010} we find the 1-$\sigma$ constraint on $\delta R_b$ is given by $-1.6 \times 10^{-4} < \delta R_b < 1.18 \times 10^{-3}$. The corresponding constraint on the $Z'$ mass and $g_{bbZ'}$ coupling are shown in \fref{FIG1a}.

\begin{figure}[t!]
\subfigure[The loop induced RHCC in top decay.]
{
\includegraphics[width=3in,height=2in,angle=0]{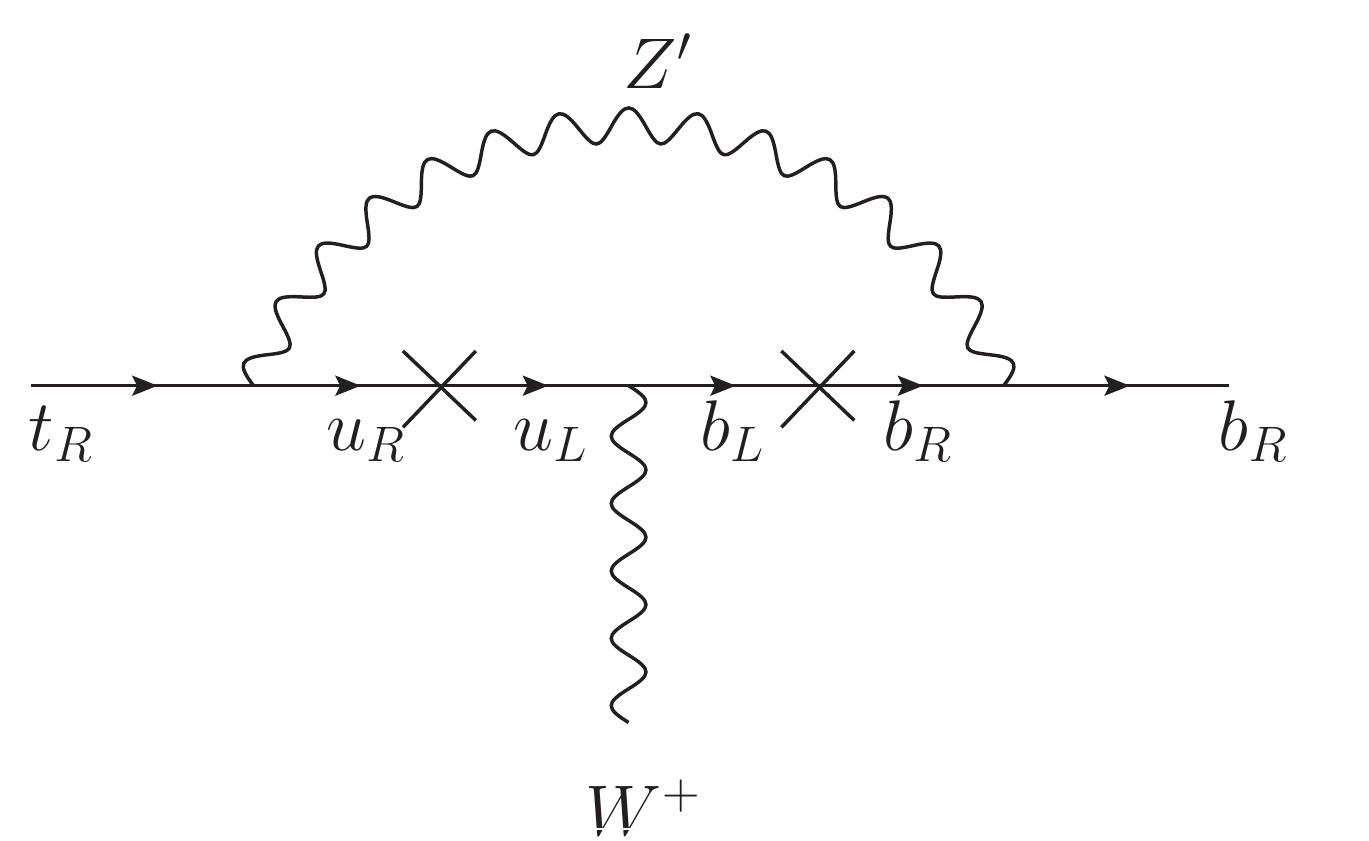}
\label{FIG2a}
}
\hspace{.25in}
\subfigure[The loop induced RHCC in $B \ra X_u \ell \nu$ decays.]
{
\includegraphics[width=3in, height=2in, angle=0]{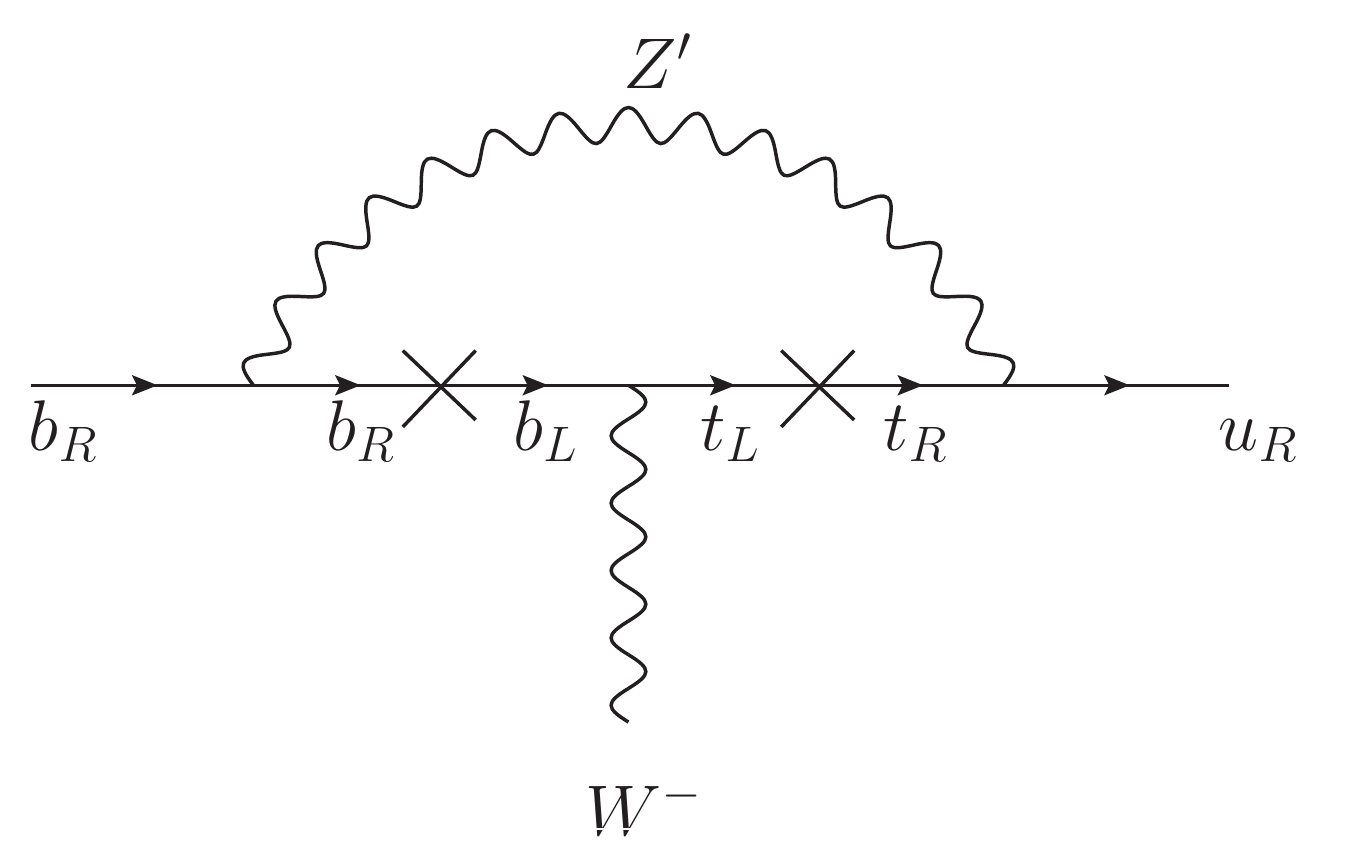}
\label{FIG2b}
}
\caption{Virtual exchange of the $Z'$ generates loop-induced RHCC in top decay and $B$ meson decay. The crosses on the internal fermion lines represent mass insertions.}
\label{FIG2}
\end{figure}

The anomalous loop-induced RHCC in the decays of the top quark and the B meson are generated by virtual $Z'$ exchange in the penguin diagrams in \fref{FIG2} \footnote{Both RHCC are calculated in the unitary gauge and are logarithmically divergent. Since there is no counter term for the RHCC these divergences must cancel. This cancellation must come from the the scalar sector which gives the $Z'$ its mass. Aside from cancelling the logarithmic divergence the scalar sector should also contribute a finite part which can be neglected by assuming that the scalar mass is large compared to all other scales in the problem. In this limit, the finite RHCC are then solely due to the finite contributions from the diagrams in \fref{FIG2a} and \fref{FIG2b}.}. The strength of the RHCC in top decay is

\begin{eqnarray}
V^R_{tb} &=& \frac{g_{utZ'} g_{bbZ'} V^{*}_{ub} }{16 \pi^2} \sqrt{x_u x_b} F_1 ( x_t, x_W ) 
\label{eq:12}
\end{eqnarray}
where $x_i = m_i^2/ M_{Z'}^2$ and the loop function, $F_1 ( x_t, x_W )$, is

\begin{eqnarray}
F_1 ( x_t, x_W ) = \int_0^1 \int_0^{1-x} dx dy \left[ \frac{ (2+ x y x_W) }{ \big( (1-x x_t) (1-x-y) - x y x_W \big) } - \ln \big( (1-x x_t) (1-x-y) - x y x_W \big) - 1 \right] . \notag \\
\label{eq:13}
\end{eqnarray} 

\begin{figure}[t!]
\centerline{\includegraphics[width=3in, height=2.75in, angle=0]{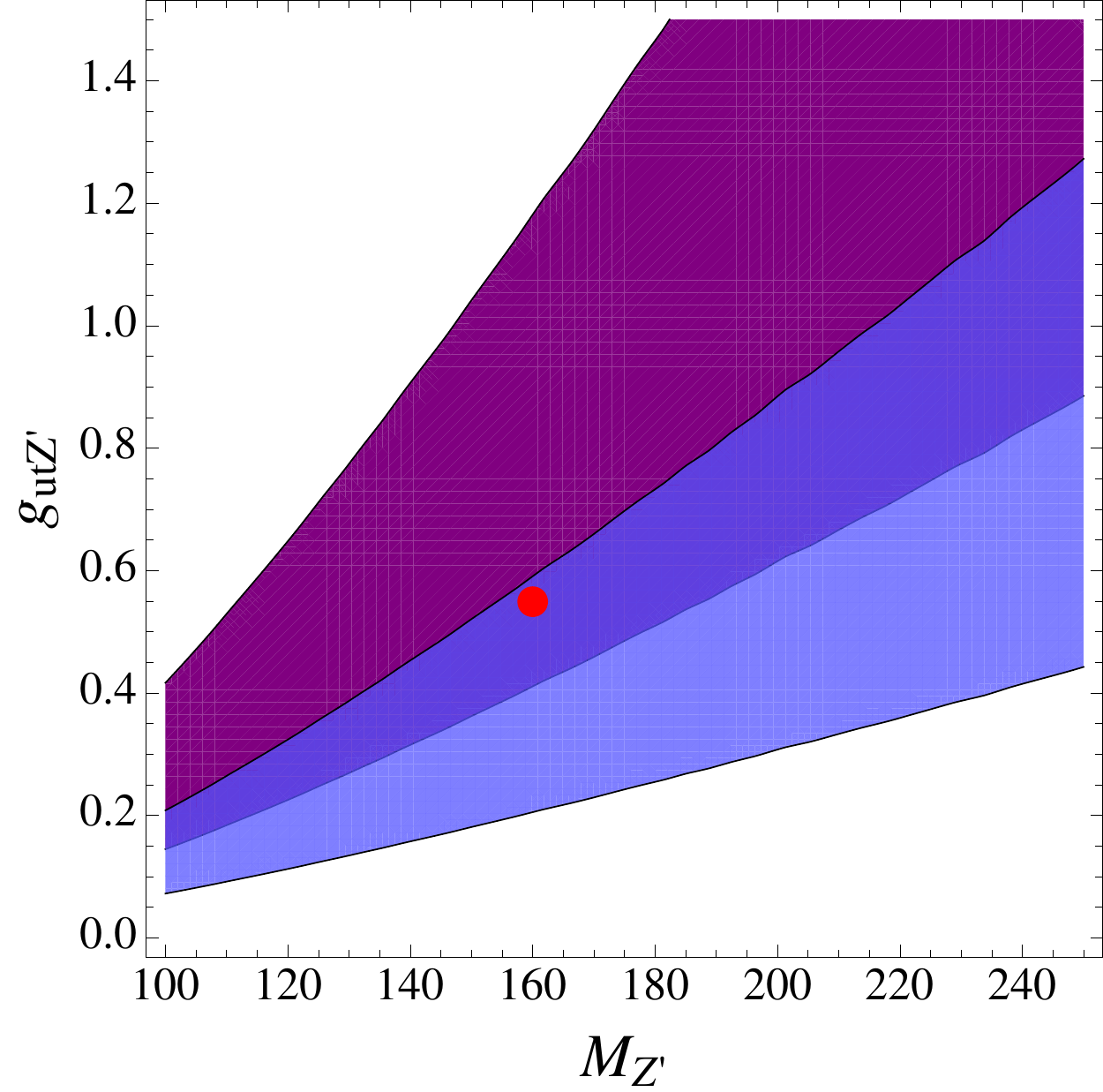}}
\caption{The $V_{ub}$ tension is removed in both the blue and the purple region. In the blue region, the flavor diagonal coupling, $g_{bbZ'}$, is set to its maximum value at each value of $M_{Z'}$ as set by constraints from $R_b$. In the purple region, we have set $g_{bbZ'}$ to half of its maximum value at each value of $M_{Z'}$. The red dot indicates the best fit point of reference \cite{Jung:2009jz} for reproducing the $t \overline{t}$ asymmetry.}
\label{FIG3}
\end{figure}

Note that the result is severely CKM as well as chiral suppressed. Even in the extreme case of choosing unreasonably large values for both couplings $g_{utZ'} = g_{bbZ'} = 5$, an extremely light value of $M_{Z'}$, e.g., $M_{Z'} = 100 \text{ GeV}$, and assuming the loop function is $\mathcal{O} (1)$ we find $V^R_{tb} \simeq 10^{-9}$ which is far below the indirect limits set by $\text{Br} ( B \ra X_s \gamma )$. 

The RHCC in $B \ra X_u \ell \nu$ decays, as shown in \fref{FIG2b}, however enjoys a chiral enhancement in comparison and is also CKM allowed. In calculating the loop function  we have neglected all external momenta so that the strength of the RHCC is given by

\begin{equation}
V^R_{ub} = \frac{g_{utZ'} g_{bbZ'}  }{ 32 \pi^2 } V_{tb} \sqrt{x_t x_b} F_2 (x_t) 
\label{eq:14}
\end{equation}
where the loop function is 

\begin{eqnarray}
F_2 ( x_t ) =  \frac{1}{2} + \frac{(4+x_t) \ln x_t}{ (x_t-1)} .
\label{eq:15}
\end{eqnarray}

Since the sign of the $g_{utZ'}$ and $g_{bbZ'}$ couplings cannot be determined we present the range of parameter space in which $1.6 \times 10^{-4} \leq \left| \text{Re} \left(  V^R_{ub}  \right) \right| \leq 4.6 \times 10^{-4}$. In order to do this we use the $R_b$ constraints to write the maximum allowed value of $g_{bbZ'}$ as a function of $M_{Z'}$ and plot the range of parameter space in which the $|V_{ub}|$ tension is alleviated in terms of $g_{utZ'}$ and $M_{Z'}$ as shown in \fref{FIG3}. We note here that the preferred region of parameter space in \fref{FIG3} contains the best fit point of reference for reproducing the $t \overline{t}$ asymmetry as well as avoiding large branching ratios for the decays $t \ra Z' u$ \cite{Jung:2009jz}. 

\begin{figure}[t!]
\subfigure[]
{
\includegraphics[width=2.5in,height=1.5in,angle=0]{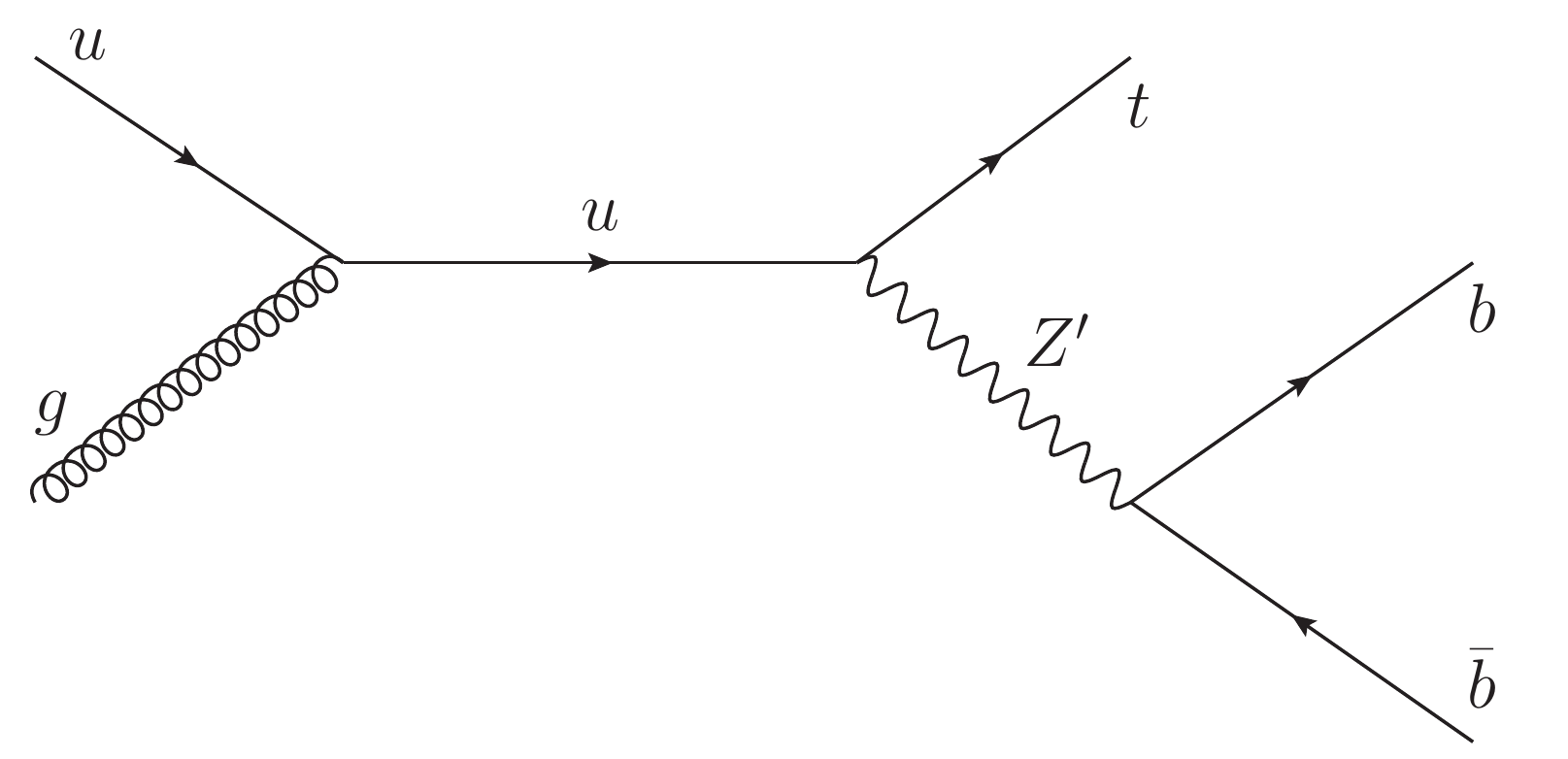}
\label{FIG4a}
}
\hspace{.5in}
\subfigure[]
{
\includegraphics[width=2.5in, height=1.5in, angle=0]{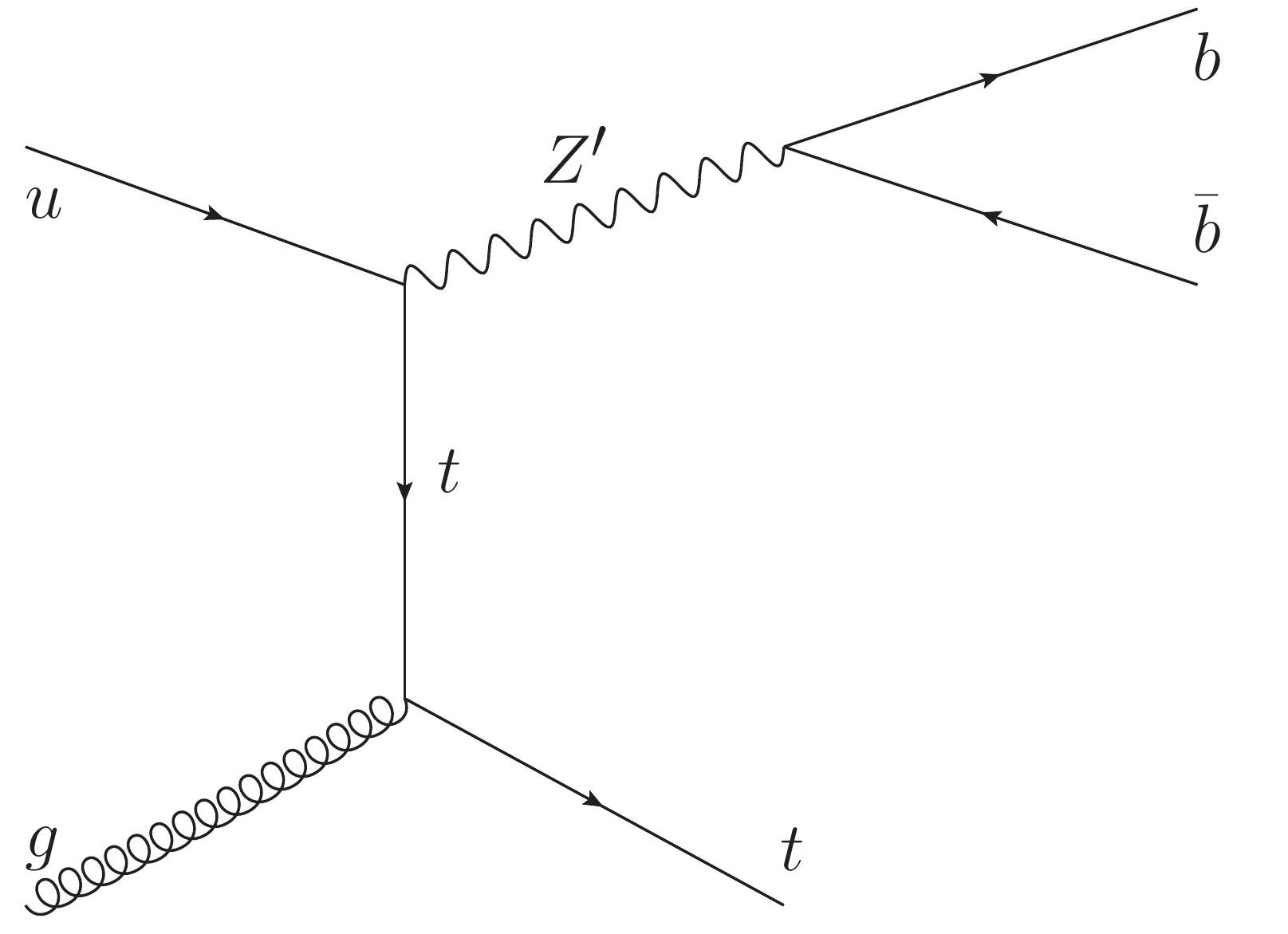}
\label{FIG4b}
}
\centering
\subfigure[]
{
\includegraphics[width=3in,height=2in,angle=0]{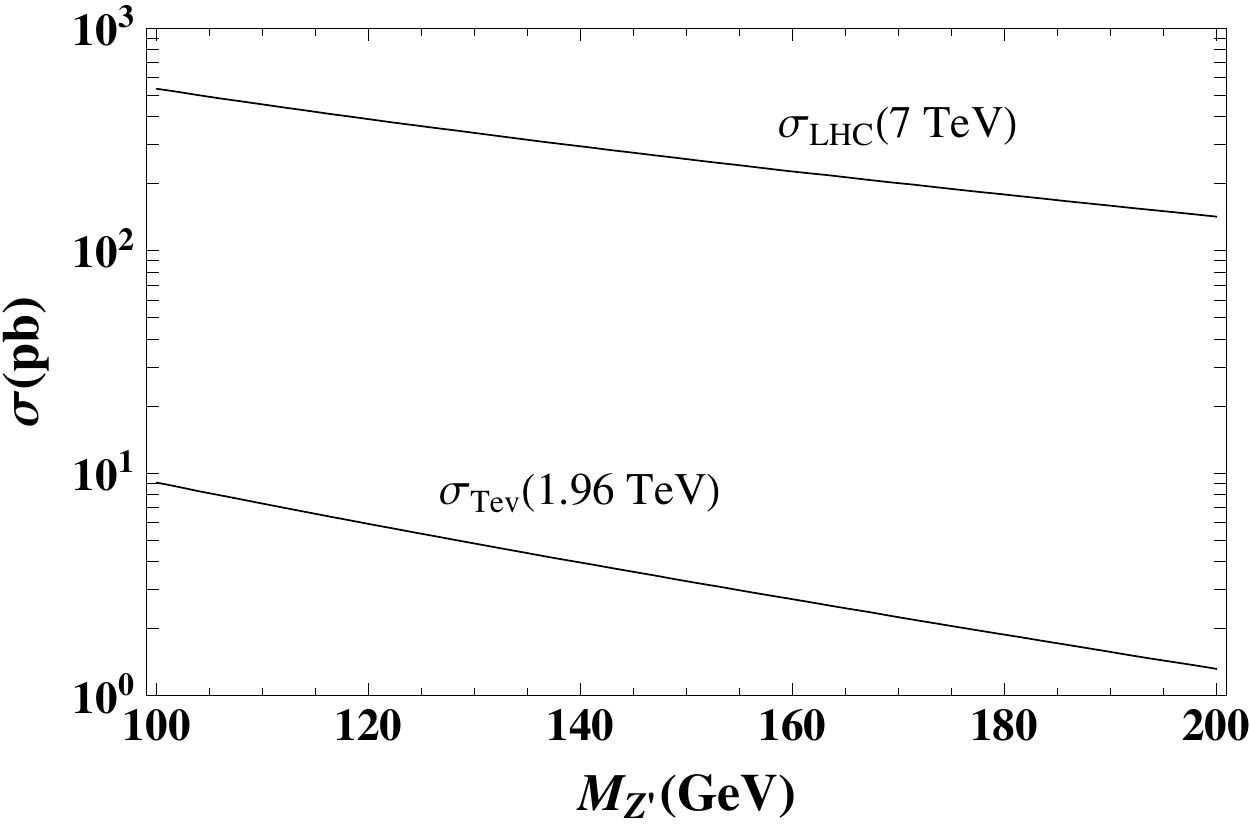}
\label{FIG4c}
}
\caption{Figures a) and b) show the diagrams for single top production in association with a $b \bar{b}$ pair from on-shell $Z'$ decay. Figure c) shows the inclusive cross sections for this signal at the LHC and the Tevatron.}
\label{FIG4}
\end{figure}
The introduction of the new flavor diagonal $b \bar{b}$ coupling also opens up the possibility of a new and interesting hadron collider signature in the form of a single top quark in association with a $b \bar{b}$ pair and no more $\slashed{E}_T$ than already associated with the top quark. The dominant production occurs when the $b \bar{b}$ pair is produced resonantly as shown in \fref{FIG4a} and \fref{FIG4b}. The mass of the $Z'$ can then be determined by examining the invariant mass distribution of the $b\bar{b}$ pair for the characteristic resonance peak. As well, as far as we can tell, there have been no searches for this particular signal. The SM background proceeds through off-shell $W^+$ boson decays or the exchange of both $W^+$ bosons and heavy flavor quarks in the t-channel and therefore has extra phase space suppression as well as CKM suppression compared to the signal. In \fref{FIG4c} we display the inclusive cross sections for the resonant case as a function of the $Z'$ mass at both the Tevatron and the LHC. The resonant production cross section is the product of the $t Z'$ production cross section and the $Z' \ra b \bar{b}$ branching ratio, $\sigma = \sigma( u g \ra t Z') Br( Z' \ra b \bar{b})$. Since the $Z'$ can only decay to $b \bar{b}$ or an off-shell top with a light jet and both the couplings involved are $\mathcal{O}(1)$, the branching ratio for $\text{Br} ( Z' \ra b \bar{b} )$ is essentially unity. Both cross sections are calculated for $g_{utZ'} = g_{bbZ'} = 1$, however, the rates for other values of $g_{utZ'}$ and $g_{bbZ'}$ are obtained from $\sigma (g_{utZ'}, g_{bbZ'}) = \sigma (g_{utZ'} = 1, g_{bbZ'} = 1) g_{utZ'}^2 g_{bbZ'}^2$.
  
The signal events are generated using MADGRAPH 5 version 1.3.29 \cite{arXiv:1106.0522} with the CTEQ6L1 parton distribution functions (pdfs) \cite{hep-ph/0201195} and the renormalization and factorization scales fixed to the MADGRAPH 5 value of the top quark mass, $m_t = 174.3 \text{ GeV}$. Although the cross sections in \fref{FIG4c} are quite large we expect the rates to reduce drastically once more relevant values of the NP couplings are used and application of kinematic cuts, detector simulation, and b-tagging efficiencies have been properly accounted for. We also briefly note here that, although the signal does not match any of the SM single top production signals, it should still be possible to use inclusive single top production cross section measurements as a constraint on the signal. As our goal for this paper is simply to demonstrate a possible viable connection between the $t \bar{t}$ FBA and the tension in the determinations of $|V_{ub}|$ we leave a more detailed analysis of the collider signatures for future work.

\section{Color Singlet Weak Doublet Scalars}
    
The possibility of a new scalar boson which couples to quarks with a non-trivial flavor structure has been studied by many authors recently \cite{BarShalom:2007pw,BarShalom:2008fq,Arhrib:2009hu,Dorsner:2009mq,Shu:2009xf,Cao:2010zb,Grinstein:2011yv,Patel:2011eh,Gresham:2011dg,Tulin:2011wi,Chang:2010et,Blum:2011fa}. In terms of the top quark FBA there are eight scalar representations capable of producing the desired interference effect with the SM however only one is a color singlet. As this representation seems to have the least amount of tension with the measurement of the $t \overline{t}$ production cross section and invariant mass distribution \cite{AguilarSaavedra:2011ug} we choose to focus our efforts on it. In what follows, our results heavily rely on the previous work of \cite{Blum:2011fa} and we therefore refer the reader to this reference for further details. 

The color singlet representation is given by $(1,2)_{-1/2}$ and can be parametrized as $\Phi = \left( \begin{array}{c} \phi^0 \\ \phi^- \end{array} \right)$ with the relevant interactions

\begin{eqnarray}
\mathcal{L} \supset X_{ij} \overline{Q}_{i_L} \Phi u_{j_R} + \tilde{X}_{ij} \overline{Q}_{i_L} \tilde{\Phi} d_{j_R} + \text{h.c.}
\label{eq:16}
\end{eqnarray}
where $i,j$ are flavor indices, $u_{i_R}$ and $d_{i_R}$ are the $SU_L(2)$ singlet quarks, and $\overline{Q}_{i_L} = ( \overline{u}_{i_L} \; \; \overline{d}_{j_L} V^*_{ij} )$ in the mass basis. In order to account for the $t \overline{t}$ asymmetry by t-channel exchange it is necessary to allow for an $\mathcal{O}(1)$ coupling between the top quark and a first generation quark. Flavor constraints imply that the first generation quark doublet should be avoided and that the $\mathcal{O}(1)$ coupling responsible for generating the asymmetry can be associated with either $u_R$ or $d_R$, dubbed ``Case II'' and ``Case VI'' respectively in \cite{Blum:2011fa}, but not both. Although a coupling to $d_R$ can provide for a large FBA, this cannot be done without running into cross section constraints so we will further focus our attention on Case II where the form of the coupling matrix, $X_{ij}$, is restricted to 

\begin{eqnarray}
X_{ij} = 
\lambda \left(
\begin{array}{ccc}
V_{ub} & 0 & 0 \\
V_{cb} & 0 & 0 \\
V_{tb} & 0 & 0
\end{array}
\right) 
\label{eq:17}
\end{eqnarray} 
with $\lambda \sim \mathcal{O}(1)$. For our purposes we also wish to introduce a flavor diagonal $b \bar{b}$ coupling in the down quark sector. This requires us to choose the following form for the $\tilde{X}_{ij}$ coupling matrix 

\begin{eqnarray}
\tilde{X}_{ij} = 
g_{bb \phi} \left(
\begin{array}{ccc}
0 & 0 & V_{ub} \\
0 & 0 & V_{cb} \\
0 & 0 & V_{tb}
\end{array}
\right) 
\label{eq:17}
\end{eqnarray} 
which also insures that no FCNC appear in the down type quark sector. The phenomenological model we work with is then 

\begin{eqnarray}
\mathcal{L} = \lambda \phi^0 \sum_{q=u,c,t} V_{qb} \bar{q}_L u_R + \lambda \phi^- \bar{b}_L u_R + g_{bb \phi} \phi^+ \sum_{q=u,c,t} V_{qb} \bar{q}_L b_R - g_{bb \phi} \phi^{* 0} \bar{b}_L b_R + \text{h.c.} 
\label{eq:18}
\end{eqnarray}

For the scalar doublet to provide a large $t \overline{t}$ FBA in the high invariant mass range, we require $|\lambda| \gtrsim 0.6$ and the mass of the neutral scalar component of the doublet to satisfy $m_0 \lesssim 130$ GeV \cite{Blum:2011fa}. Considering LEPII searches we also restrict ourselves to a lower limit of $m_0 \geq 100$ GeV. We note briefly the existence of new terms in our model which alter $t \bar{t}$ production. However, since the only process which is altered by the added terms is $b \bar{b} \ra t \bar{t}$, which is highly parton luminosity suppressed anyway, these alterations have no significant effect on the total production cross section or the FBA at the Tevatron.

\begin{figure}[t!]
\centerline{\includegraphics[width=4in,height=3in,angle=0]{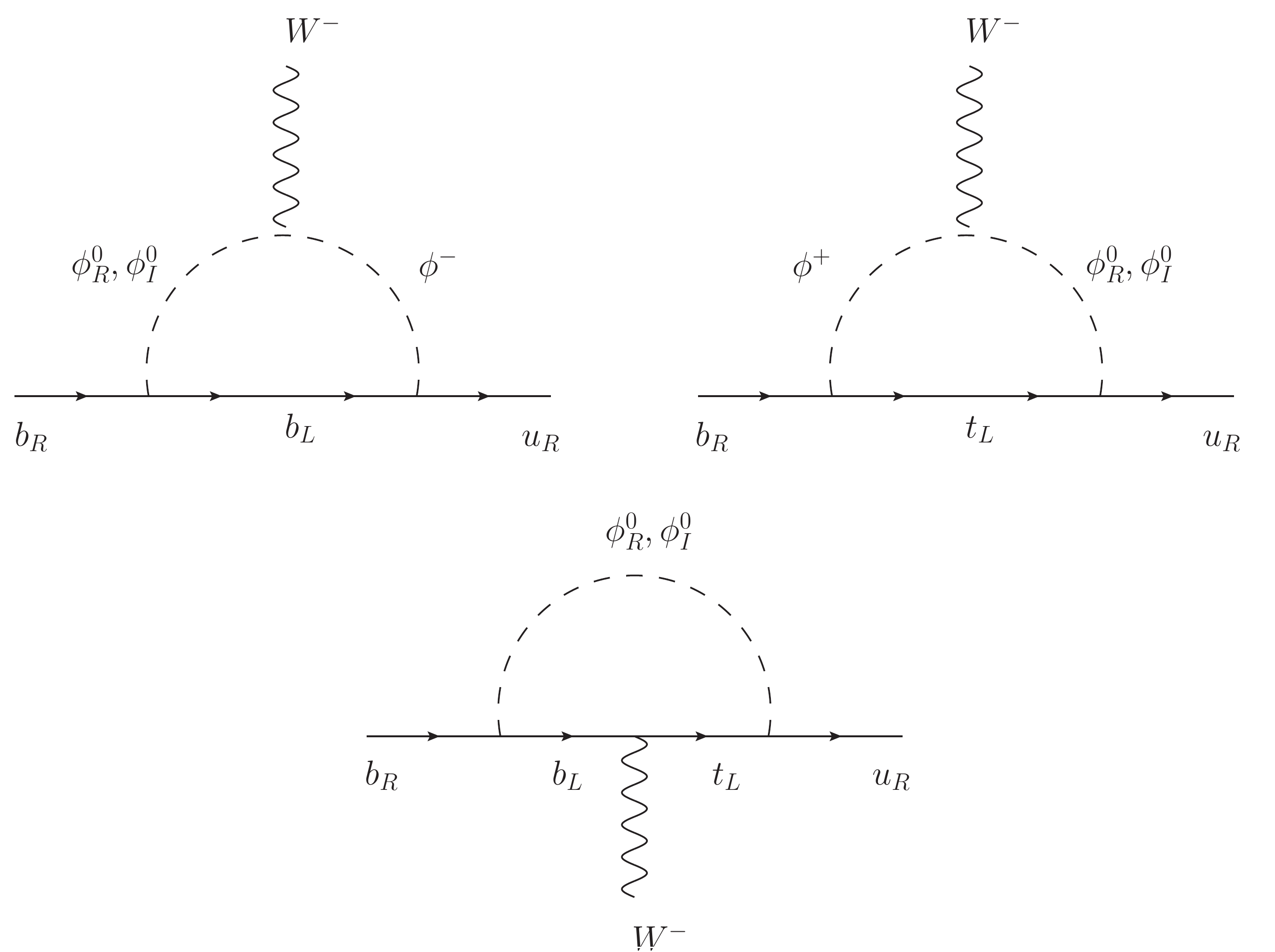}}
\caption{The six CKM allowed diagrams which generate the RHCC in $B \ra X_u \ell \nu$ decays.}
\label{FIG5}
\end{figure}

Since flavor constraints, in particular $K^0$-$\overline{K}^0$ and $D^0$-$\overline{D}^0$ mixing constraints, disfavor a large coupling to the right-handed top quark there are no RHCC in top decays and the model remains free from indirect $B \ra X_s \gamma$ constraints. As well, because we do not add a coupling to the $s$ quark, the model also remains free from direct $B \ra X_s \gamma$ constraints. The neutral scalar is complex and is parametrized as $\phi^0 = \frac{1}{\sqrt{2}} ( \phi^0_R + i \phi^0_I )$, leading to both scalar and pseudoscalar current diagrams contributing to the RHCC in $B \ra X_u \ell \nu$ decays. Of these contributions only six of the diagrams are CKM allowed, which we show in \fref{FIG5}. Other contributions are neglected as they are CKM suppressed and do not greatly affect our results. The strength of the RHCC depends on the normalized masses of the neutral and charged scalars, $x_i = m_i^2/M_W^2$, where $i = R, I, +$ denotes the real, pseudo, and charged scalar masses. We find that the RHCC vanishes for zero mass splitting between the real and pseudo scalars and instead prefers a large splitting and a small charged scalar mass. The result is finite, with exact cancellation of logarithmic divergences occurring between diagrams with scalar and pseudoscalar currents, and is given by 

\begin{equation}
V^R_{ub} = - \frac{g_{bbZ'} \lambda}{64 \pi^2} \bigg( 2 A_0 ( x_+, x_I, x_R ) - A_0 (x_t, x_I, x_R ) + B_0 (x_+, x_R, x_t ) - B_0 ( x_+, x_I, x_t ) \bigg) 
\label{eq:19}
\end{equation}
with loop functions

\begin{equation}
A_0 ( x_a, x_b, x_c ) = \frac{x_a x_b \ln \bigg( \ds \frac{x_a}{x_b} \bigg) + x_b x_c \ln \bigg( \ds \frac{x_b}{x_c} \bigg) + x_c x_a \ln \bigg( \ds \frac{x_c}{x_a} \bigg)  }{(x_c - x_a) (x_a - x_b) }  \notag
\label{eq:20}
\end{equation}

\begin{equation}
B_0 (x_a, x_b, x_c) = - \frac{x_t}{(x_b - x_c)} A_0 ( x_a, x_b, x_c )
\label{eq:21}
\end{equation}

Constraints from the electroweak $T$ parameter restrict the splitting between the neutral and charged scalar masses to be $|m_+ - m_{R_0}| \lesssim 110$ GeV \cite{Blum:2011fa}. If both the scalar and pseudoscalar masses are bound such that $100 \text{ GeV} \leq m_{R_0}, m_{I_0} \leq 130 \text{ GeV}$ then we find that the loop function peaks for $m_R = 130$ GeV and $m_I = m_+ = 100$ GeV. 

\begin{figure}[t!]
\begin{center}
\centerline{\includegraphics[width=3in,height=3in,angle=0]{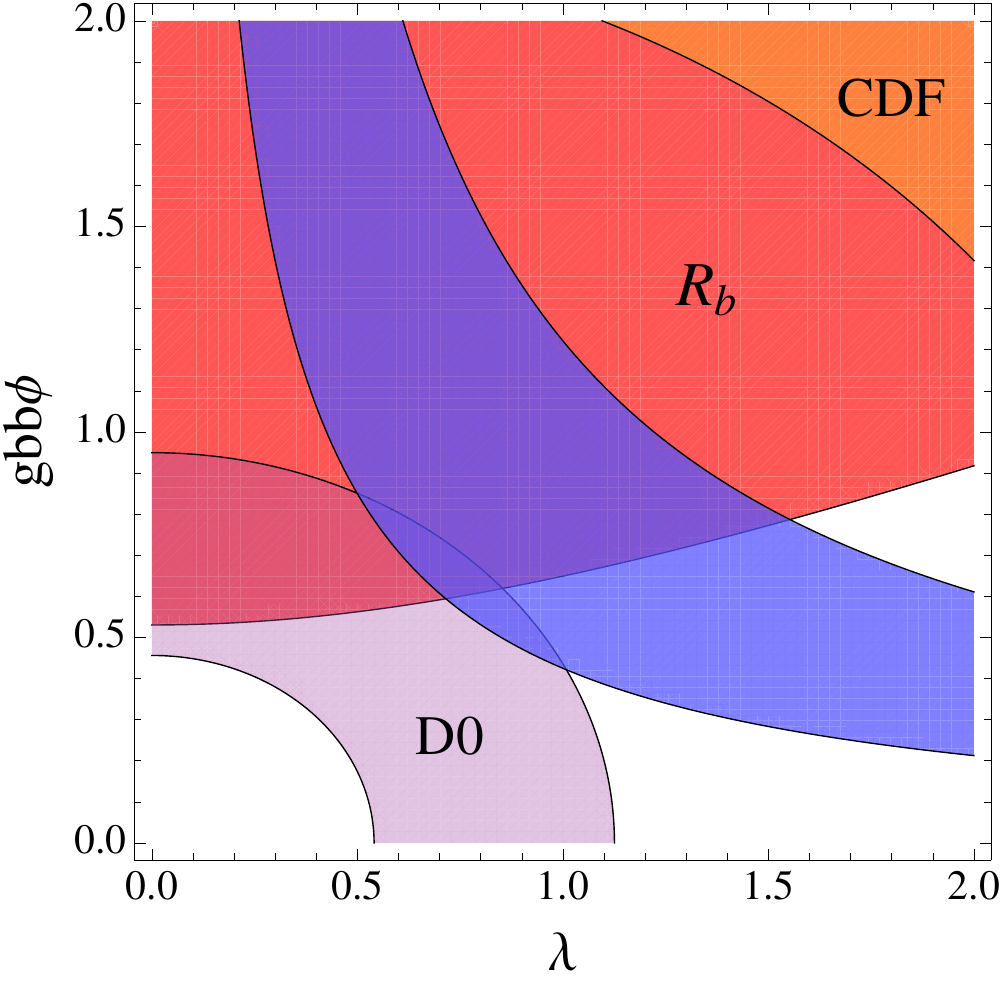}}
\end{center}
\caption{The red and yellow regions are excluded by the bounds on $R_b$ and the top quark width from the CDF collaboration respectively. The purple and blue regions are the preferred regions of parameter space for the indirect determination of the top quark width from the D0 collaboration and the alleviation of the $|V_{ub}|$ tension using $m_{R_0} = 130$ GeV and $m_{I_0} = m_\pm = 100$ GeV.}
\label{FIG6}
\end{figure}

The authors of \cite{Blum:2011fa} found that the parameter space of ``case II'' remained unconstrained by $R_b$ bounds however, in light of the new flavor diagonal coupling $g_{bb \phi}$, these bounds must be revisited anew. We find seven diagrams, all topologically similar to the RHCC diagrams, contributing to $R_b$. The shifts in the $Zb\overline{b}$ couplings are also finite, with the logarithmic divergences cancelling in the same pattern as in the RHCC, and have the following form

\begin{eqnarray}
\delta g_{b_L} = \bigg( \frac{g_{bb \phi}}{4 \pi} \bigg)^2 C_0 (z_R, z_I ) + \bigg( \frac{\lambda}{4 \pi} \bigg)^2 D_0 (z_+ ) \; \; \; \; \; \; \; \; \; \; \; \; \delta g_{b_R} = \bigg( \frac{g_{bb \phi}}{4 \pi} \bigg)^2 E_0 ( z_R, z_I, z_+ )
\end{eqnarray}

where $z_i = m_i^2/M_Z^2$ with $i = +, R, I$ and the above loop functions taking the values  

\begin{eqnarray}
C_0 (z_R, z_I ) = 0.024 \; \; \; \; \; \; \; \; \; \; \; \; D_0 (z_+ ) = -0.016 \; \; \; \; \; \; \; \; \; \; \; \; E_0 ( z_R, z_I, z_+ ) = -0.5
\end{eqnarray}
when the $V^R_{ub}$ preferred values for the masses are chosen. We find that there exists a region of parameter space which remains free from $R_b$ constraints and also coincides with the preferred parameter space for the removal of the $|V_{ub}|$ tension which we present in \fref{FIG6}. Although the opposite signs and similar magnitudes of the $C_0(z_R, z_I)$ and $D_0(z_+)$ loop functions imply the existence of a non-trivial region of cancellation in \fref{FIG6} this region does not present itself within the perturbative regime of the couplings. We also note here that the region in which the $|V_{ub}|$ tension is alleviated corresponds to the region in which a large top quark FBA can be generated.

Due to the existence of new couplings between the top and bottom quarks and the charged and neutral scalars, whose masses are set at $\mathcal{O} (100) $ GeV to accommodate a large RHCC, multiple new decay modes for the top quark are opened up. Defining 

\begin{eqnarray}
\Gamma ( g, m) = \frac{g^2 m_t} {32 \pi} \bigg( 1 - \frac{m^2}{m_t^2} \bigg)^2  
\end{eqnarray}
the total width of the top quark, at LO and excluding CKM suppressed SM decay modes, becomes

\begin{align}
\Gamma_t = \Gamma_t^{SM} + \Gamma \bigg( \frac{\lambda |V_{tb}|}{\sqrt{2}}, m_R \bigg) +  \Gamma \bigg( \frac{\lambda |V_{tb}|}{\sqrt{2}}, m_I \bigg) +  \Gamma \bigg( g_{bb\phi} |V_{tb}| , m_+ \bigg) .
\end{align}
where we will take $\Gamma_t^{SM} = 1.3$ GeV. The CDF collaboration has performed a direct measurement of the top quark width using $t \bar{t}$ events produced in $p \bar{p}$ collisions at the Tevatron and set an upper limit of $\Gamma_t < 7.6$ GeV  \cite{Aaltonen:2010ea}. The D0 collaboration has also recently released an indirect determination of the total width of the top quark by combining the measurements of the single top t-channel production cross section and the branching ratio $Br ( t \ra W b )$ as measured in $t \bar{t}$ events leading to the result $\Gamma_t = 2.00^{+ 0.47}_{-0.43}$ \cite{Abazov:2012vd}. Using $|V_{tb}| = 1$ and $m_t = 174.3$ GeV and the RHCC preferred values of the scalar masses the parameter regions which satisfy the top width bounds are also presented in \fref{FIG6}.

The D0 collaboration has also performed a model independent measurement of the t-channel single top quark production cross section $\sigma ( p \bar{p} \ra t b q + X ) = 2.90 \pm 0.59$ pb \cite{Abazov:2011rz} where $ t b q = t \bar{b} q + \bar{t} b q$, which is in good agreement with the SM prediction \cite{Kidonakis:2006bu}. At the Tevatron single top quarks are produced in the t-channel by exchange of a $W$ boson between a light quark and a b quark from the sea. Since the sea is flavor symmetric the presence of a b quark in the initial state guarantees the presence of an anti b quark in the final state. In the phenomenological model in which we work, Eq. \ref{eq:18}, single top production in the t-channel can arise from 2 $\ra$ 2 processes or 2 $\ra$ 3 processes. In the 2 $\ra$ 2 processes, the signal could be generated by either neutral or charged scalar exchange. For neutral scalar exchange, $u q \ra t q'$, either neither $q$ or $q'$ are b quarks, in which case there is no b quark in the final state, or both $q$ and $q'$ are b quarks, in which case there are too many b quarks in the final state. For charged scalar exchange, in order to get a single top in the final state a b quark needs to be extracted from the sea leading to the presence of a b quark in the final state. However, since all charged scalar interactions involve a b quark, the presence of too many b quarks in the end state is guaranteed. The same issue occurs in all 2 $\ra$ 3 processes which leads us to conclude that the constraint does not directly apply.

As mentioned before in the context of the leptophobic $Z'$ model, there exist tight constraints associated with the non-observation of a significant amount of same-sign top quark events at both the Tevatron and the LHC. In the phenomenological model which we use the only production mode is $uu \ra tt$ via t and u channel exchange of the real and pseudo scalars. Although this limit strongly constrains the leptophobic $Z'$ we find that the color singlet weak doublet scalar model is not strongly bound by these results due to a cancellation between the real and pseudo scalar contributions which is exact in the limit of degenerate scalar masses. This cancellation occurs because the neutral scalar is, unlike the $Z'$, complex and, for this reason, we find that the model remains free from same sign top constraints. 

We also note briefly that dijet constraints are not relevant here due to a combination of the cancellation described above as well as CKM and parton luminosity suppression in conjunction with large SM backgrounds.

As before, the introduction of the new $b \bar{b}$ coupling also leads to the new signal discussed earlier, i.e; single top with a $b \bar{b}$ pair from an on-shell decay of a new particle and no more $\slashed{E}_T$ than already associated with the top quark. In this case, the new particles are the real and pseudo scalars and the diagrams are the same as in \fref{FIG4a} and \fref{FIG4b} except with $\phi^0_R, \phi^0_I$ in place of the $Z'$. The total cross section is given by the sum $\sigma = \sum_{i=R,I} \sigma ( u g \ra t \phi^0_i ) Br ( \phi^0_i \ra b \bar{b} )$ and the branching ratios are, again, essentially unity due to CKM suppression of all other decay modes. For illustrative purposes we choose the RHCC preferred values for the scalar masses which leads to $\sigma = \lambda^2 g_{bb\phi}^2 43.8$ pb at the LHC for $\sqrt{s} = 7$ TeV and $\sigma = \lambda^2 g_{bb\phi}^2 0.60$ pb at the Tevatron with $\sqrt{s} = 1.96$ TeV. The signal events were generated with MADGRAPH 5 and the CTEQ6L1 pdfs with both the renormalization and factorization scales set to the top mass. Again, we postpone a more involved analysis of the signal for future work.

\section{Flavor Violating Squark Mass Insertions in the MSSM}

In the minimal supersymmetric standard model (MSSM) the largest contribution to the loop-induced RHCC comes from gluino-squark loops. In this case, the flavor violation resides in the squark sector as opposed to the quark sector and is due to the fact that the bi-unitary rotations which diagonalize the SM quark mass terms do not necessarily diagonalize the squark mass terms or the Higgs-squark-squark couplings. As well as flavor violation these rotations also induce chiral mixing between squark states after the Higgs doublets acquire their respect vevs, generally leading to non-diagonal 6$\times$6 mass matrices for both the up and down type squarks. In \cite{Crivellin:2008mq} bounds on the flavor off-diagonal elements of these mass matrices were derived by requiring that no large cancellations between tree level SM CKM elements and SUSY loop corrections occur. Since the SM CKM elements reside completely in the left-handed quark sector, this leaves mixing associated with the right-handed quark sector relatively unconstrained. In \cite{Crivellin:2009sd} this freedom was exploited to show that a loop-induced RHCC generated within the MSSM from a finite gluino-squark loop can alleviate the $|V_{ub}|$ tension. We revisit this calculation here, emphasizing the importance of the particular flavor violating chiral squark mass insertions involved. 

\begin{figure}[t!]
\subfigure[The dominant gluino-squark loop contribution to the RHCC associated with top decay.]
{\includegraphics[width=3in,height=2in,angle=0]{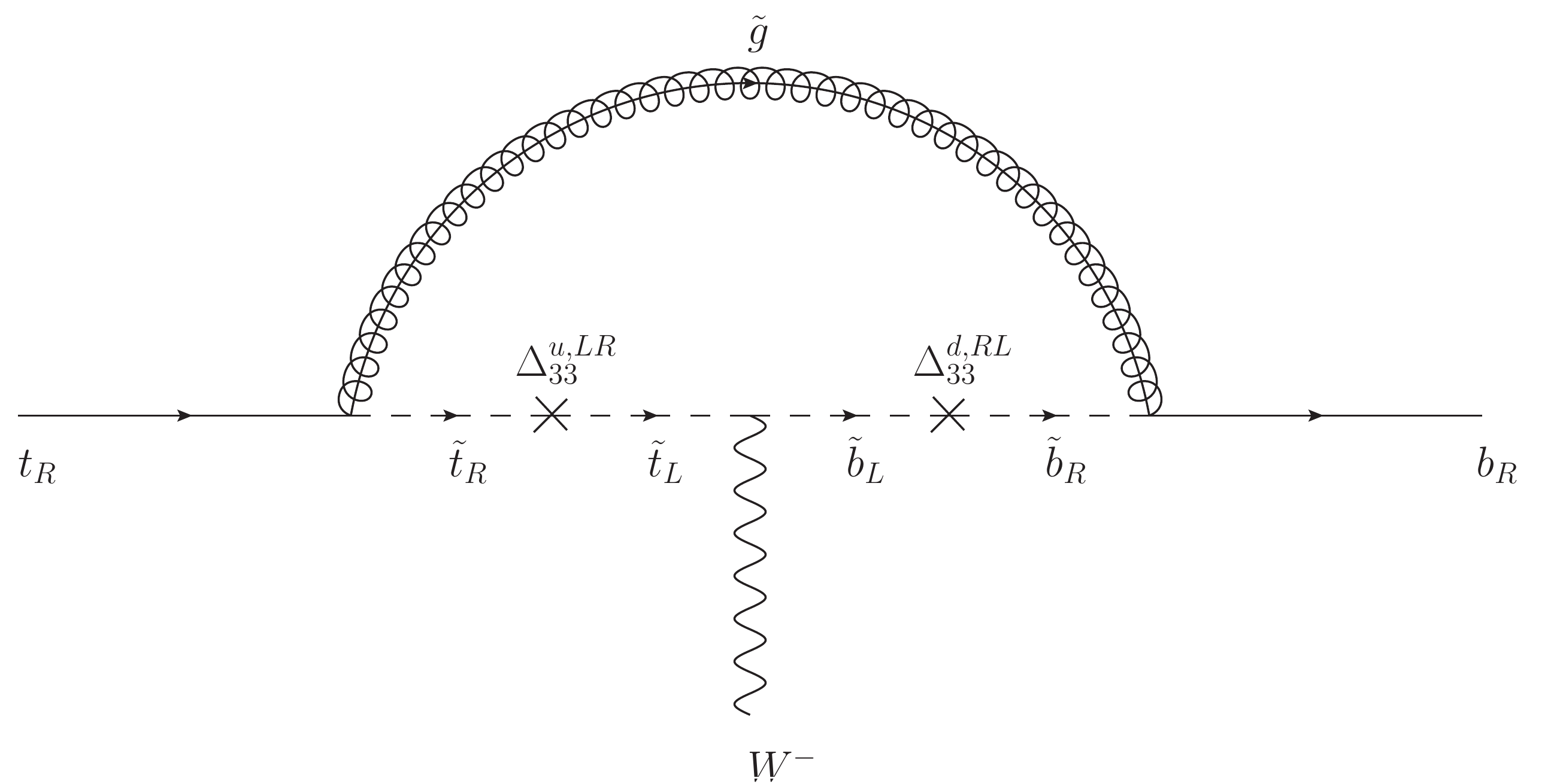}
\label{FIG6a}}
\hspace{.2in}
\subfigure[The dominant gluino-squark loop contribution to the RHCC associated with $B \ra X_u \ell \nu$ decays.]
{\includegraphics[width=3in,height=2in,angle=0]{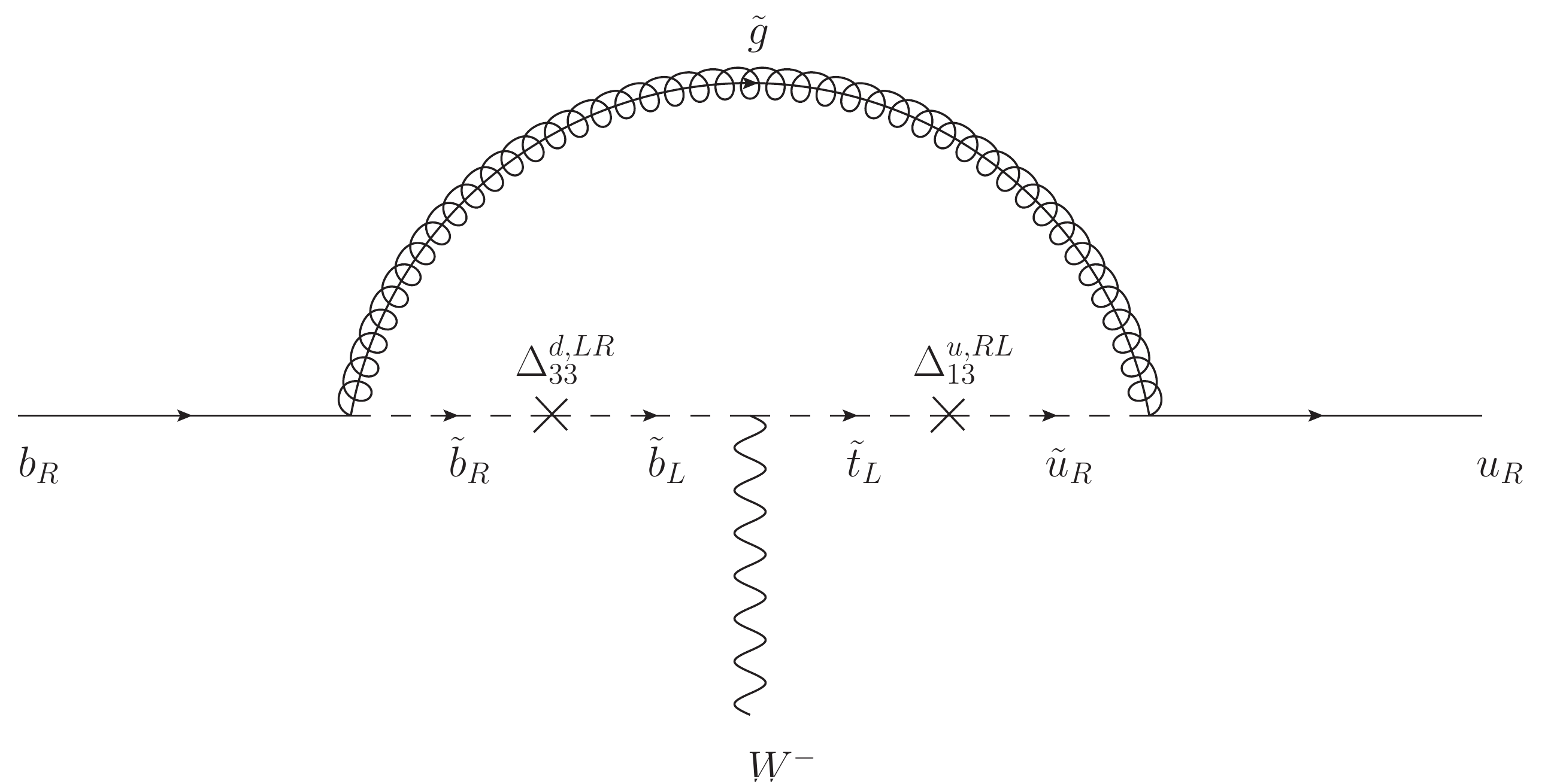}
\label{FIG6b}}
\caption{}
\end{figure}

In \fref{FIG6a} and \fref{FIG6b} we present the dominant gluino-squark loop diagrams which generate the RHCC in top and $B \ra X_u \ell \nu$ decays in the MSSM. We have chosen to work in the super-CKM basis and treat the flavor off-diagonal elements of the squark mass matrices, written as $\Delta^{\tilde{q},XY}_{ij}$ in the notation of \cite{Crivellin:2008mq}, as perturbations. We also define the dimensionless quantities $\delta^{XY}_{ij} = \Delta^{XY}_{ij} / m^2_{\tilde{q}}$ and take all squark masses to be equal to the average squark mass $m^2_{\tilde{q}} = \ds \frac{1}{6} \sum_s m^2_{\tilde{q}_s}$.

The dominant RHCC in top decay involves only chiral mixing elements in the squark mass matrix which are not constrained by flavor observables. Our result for the RHCC in top decay is

\begin{eqnarray}
V^R_{tb} = \frac{4 \alpha_s}{3 \pi} V_{tb} \delta^{d,RL}_{33} \delta^{u,LR}_{33} F_3 (x_{\tilde{g}}, x_t, x_W ) 
\label{eq:21}
\end{eqnarray}
where $x_i = m_i^2 / m^2_{\tilde{q}}$ and the loop function is given by 

\begin{eqnarray}
F_3  (x_{\tilde{g}}, x_t, x_W ) = \int_0^1 \int_0^{1-x} dx dy \frac{x y}{\Delta^2} \left( \frac{x y x_t}{\Delta} - 1/2 \right)  \notag \\
\label{eq:22}
\end{eqnarray}
with $\Delta = x + y + ( x_{ \tilde{g}} - x x_t ) ( 1 -x -y) - x y x_W$. Choosing $m_{\tilde{g}} = m_{\tilde{q}} = M_{SUSY} = 2$ TeV and requiring that $\delta^{d,RL}_{33}, \delta^{u,LR}_{33} \leq 1$ we find that $V^R_{tb}$ remains free from the indirect $\text{Br}( B \ra X_s \gamma )$ constraints.

The dominant contribution to the RHCC associated with $B \ra X_u \ell \nu$ decays instead involves the normalized flavor violating mass insertion $\delta^{u, RL}_{13}$ as all other contributions are CKM suppressed. We note here that $\delta^{u,RL}_{13}$ remains unconstrained by the naturalness bounds in \cite{Crivellin:2008mq} and, as far as we know, is not constrained at all. We find the RHCC in $B \ra X_u \ell \nu$ decays is given by

\begin{eqnarray}
V^R_{ub} = - \frac{\alpha_s}{18 \pi} V_{tb} \delta^{d,LR}_{33} \delta^{u,RL}_{13} \left[ \frac{ ( x_{\tilde{g}} - 1) ( x_{\tilde{g}} ( 5 + 2 x_{\tilde{g}}) - 1) - 6 x_{\tilde{g}}^2 \ln (x_{\tilde{g}})}{ ( x_{\tilde{g}} - 1)^4} \right] .
\label{eq:23}
\end{eqnarray}

\begin{figure}[t!]
\subfigure[The blue region removes the $|V_{ub}|$ tension via RHCC generated by gluino-squark loops involving flavor violating squark mass insertions in the MSSM.]
{\includegraphics[width=3in,height=3in,angle=0]{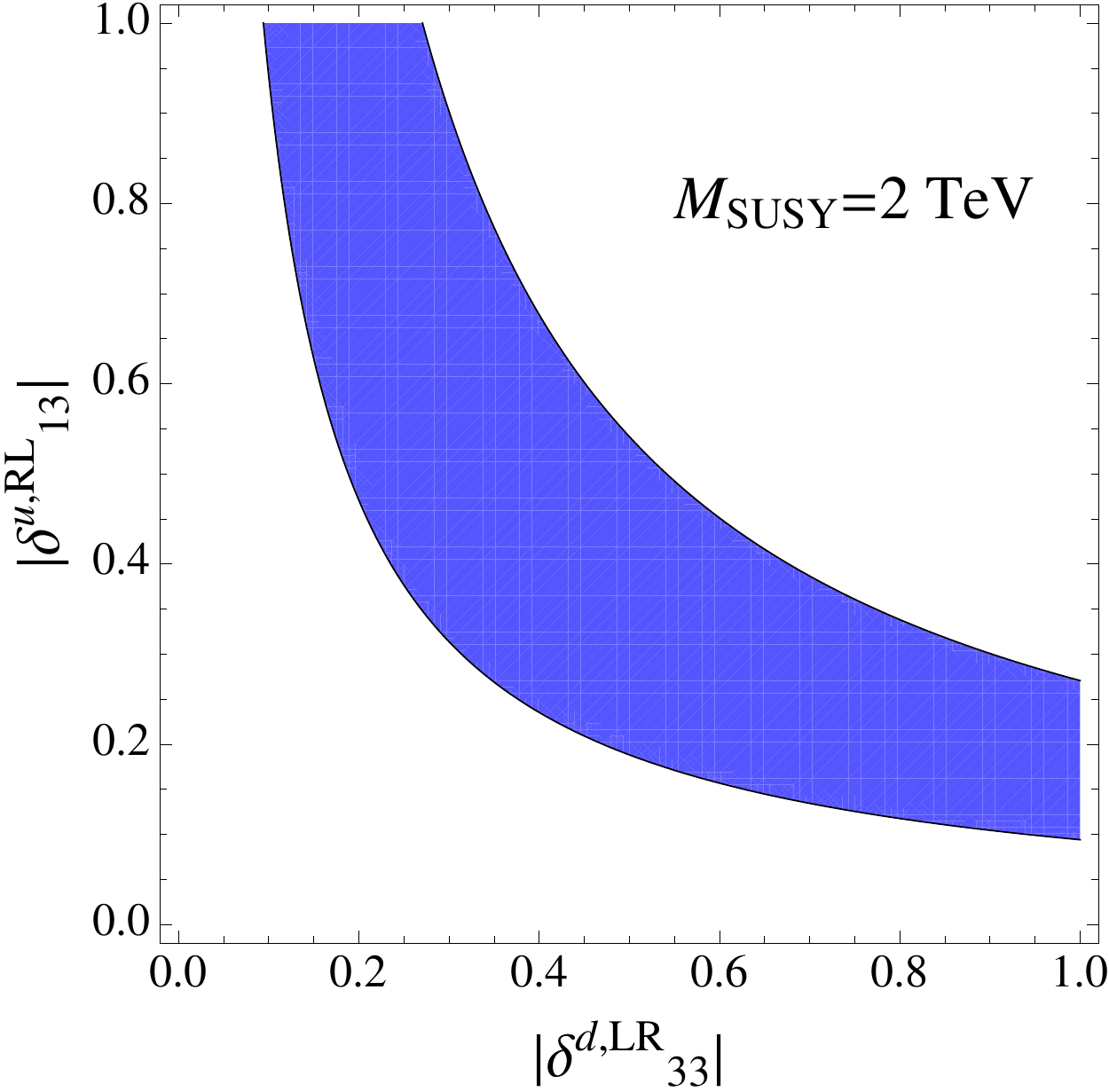}
\label{FIG7a}}
\hspace{.3in}
\subfigure[The blue region removes the $|V_{ub}|$ tension via RHCC generated by the Isidori-Kamenik model using $Y_b = 0.38$.]
{\includegraphics[width=3in,height=3in,angle=0]{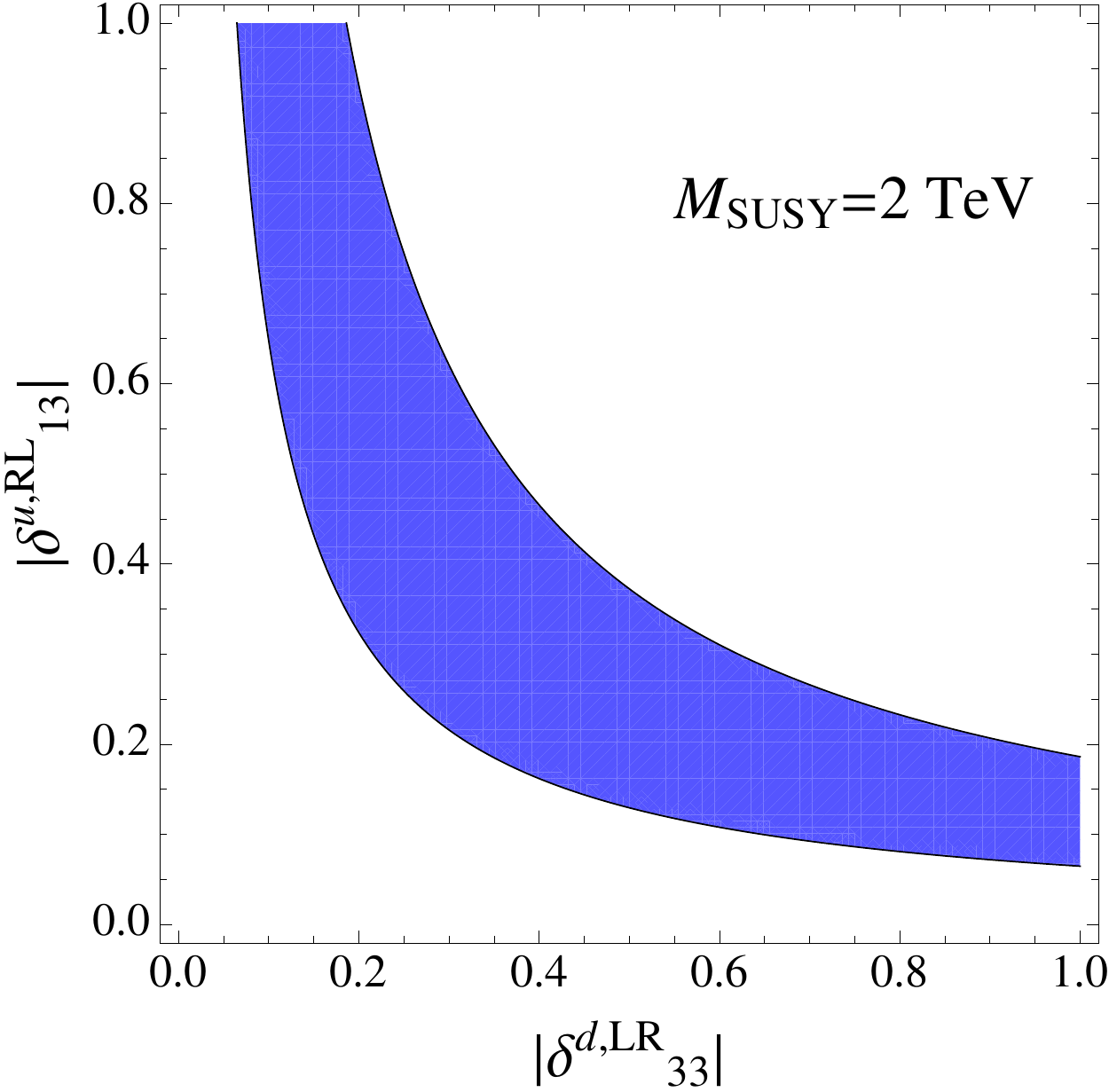}
\label{FIG7b}}
\caption{}
\end{figure}

The preferred parameter space for the alleviation of the $|V_{ub}|$ problem is shown in \fref{FIG7a} for $M_{SUSY} = m_{\tilde{g}} = m_{\tilde{q}} = 2$ TeV. Our results agree in general with those of \cite{Crivellin:2009sd}.

If one instead adopts the hypothesis of flavor-blindness then both squark mass matrices are diagonal. We find that without the aid of the flavor violating mass insertions the size of the RHCC in $B \ra X_u \ell \nu$ decays is drastically reduced. It is well known that the MSSM alone cannot address the top quark FBA at the tree level. \\

To address the FBA in a SUSY scenario, one needs to add explicit flavor changing couplings in the up quark sector. Such a model is presented in \cite{Isidori:2011dp} which we dub the Isidori-Kamenik model. The model can be thought of as an extension of the MSSM to include a new light $SU(2)_L \times U_Y(1)$ singlet majoranna particle, $\chi^0$, which interacts with the right-handed stop and each of the up type SM quarks. For our purposes we add to the Isidori-Kamenik model a flavor diagonal $\overline{b}_R \tilde{b}_R \chi^0$ coupling so that the lagrangian can be written as

\begin{equation}
\mathcal{L} = \mathcal{L}_{MSSM} + |D_\mu \tr |^2 - m^2_{\tr} \tr^\dagger \tr + \overline{\chi^0} i \gamma_\mu D^\mu \chi^0 - m_\chi \overline{\chi^0}_c \chi^0 + \sum_{q=u,c,t} \tilde{Y}_q \overline{q}_R \tr \chi^0 + Y_b \overline{b}_R \tilde{b}_R \chi^0 + \text{h.c.}
\label{eq:24}
\end{equation}

In order to generate a large FBA and maintain a small impact on cross section measurements the parameter space must be restricted such that $200 \lesssim m_{\tr} \lesssim 205$ GeV, $\tilde{Y}_u \gtrsim 1.5$, $\tilde{Y}_t \sim 4$, and $m_\chi \sim 2$ GeV. We further conservatively choose our sbottom masses such that $m_{\tilde{b}} = M_{SUSY} = 2$ TeV to be consistent with LHC SUSY particle searches. The diagrams contributing to top and $B \ra X_u \ell \nu$ decays are the same as in the MSSM with the gluino replaced by $\chi^0$. Using the prescribed parameter space we find the RHCC in top and $B \ra X_u \ell \nu$ decays can be written as

\begin{eqnarray}
V^R_{tb} = Y_b \delta^{d,RL}_{33} \delta^{u,LR}_{33} 1.9 \times 10^{-2} \notag \\ \notag \\
V^R_{ub} = Y_b \delta^{d,LR}_{33} \delta^{u,RL}_{13} 6.5 \times 10^{-3} .
\label{eq:25}
\end{eqnarray}

Taking the mass insertions in $V^R_{tb}$ to be both equal to 1 we require $Y_b \lesssim 0.38$ to satisfy $\text{Br} ( B \ra X_s \gamma )$ constraints. Using the maximum value of $Y_b$ then allows us to plot the parameter space for which the $|V_{ub}|$ tension is removed in \fref{FIG7b}. 

Since the stop masses are heavier than the top quark mass there are no new top decay modes opened up. It has also been noted that, considering the small size of $\tilde{t}_R \tilde{t}_R^\dagger$ production compared to the dominant SM backgrounds, constraints from single top production are not expected to be particularly strong \cite{Isidori:2011dp}. Although same sign top events can occur through t-channel exchange of the Majorana particle $\chi^0$ and subsequent $\tilde{t}_R$ decays this is severely suppressed by the ratio $(m_{\chi} / m_{\tilde{t}_R} )^2$, leading to cross sections well below current bounds.

With the introduction of the flavor diagonal $\overline{b}_R \tilde{b}_R \chi^0$ coupling, one might question whether this can lead to non-trivial constraints from the new decay mode of the Upsilon meson, $\Upsilon (1 S) \ra \chi^0 \chi^0$, via an intermediate sbottom squark in the $t$-channel. The decay rate can be estimated as $\sim Y_b^4 m_{\Upsilon}^5 / 2 \pi M_{SUSY}^4 = 1.6 \times 10^{-11}$ GeV, using $Y_b = 0.38$, whereas the total Upsilon width is $\Gamma_{\Upsilon} = 54 \times 10^{-6}$ GeV \cite{Nakamura:2010}. Since the $\chi^0$'s are light and SM singlets they would contribute to the $\text{Br}(\Upsilon \ra \text{invisible})$ branching ratio, for which the current limits are $\text{Br}(\Upsilon \ra \text{invisible}) < 3 \times 10^{-4}$ at 90\% C.L. \cite{Nakamura:2010}. We find that $\text{Br} ( \Upsilon \ra \chi^0 \chi^0 ) \sim 3 \times 10^{-7}$, which lies well within the current bounds.

Finally, we also briefly note here that the Isidori-Kamenik model does not have the $b \bar{b}$ type coupling which would lend itself to the new single top with $b \bar{b}$ pair signal discussed in the previous two sections. 

\section{Conclusions}

Recent measurements of the FBA of the top quark in $t \overline{t}$ production at the Tevatron have inspired many new physics models which incorporate a new flavor changing coupling involving the first and third generation quarks as well as some new mediator which can be scalar or vector in character. It was also shown recently that the effect of right-handed couplings of SM quarks to the W boson, or right handed charge currents, can alleviate the tension between the determinations of $|V_{ub}|$ from exclusive and inclusive measurements as well as from the rare leptonic $B \ra \tau \nu$ decays. We have studied the effect of such new flavor changing couplings on the size of loop-induced right-handed charge currents, making the connection between the $t \bar{t}$ asymmetry and the $|V_{ub}|$ tension by adding to each of the benchmark models considered a flavor diagonal coupling to $b \bar{b}$. The effect of such a flavor changing coupling is to simultaneously enhance right-handed charge currents in $B \ra X_u \ell \nu$ decays, which allows for the removal of the $|V_{ub}|$ tension, and suppress them in non-standard top decays which frees the given model from existing indirect yet strong $\text{Br} ( B \ra X_s \gamma )$ constraints. This situation comes about for various reasons: for a new scalar (vector) mediator the flavor changing coupling provides CKM (as well as chiral) enhancements in $B \ra X_u \ell \nu$ decays while concurrently providing CKM (as well as chiral) suppression in non-standard top decays. We have also revisited and confirmed the possibility that the right-handed charge current generated by flavor violating squark mass insertions in the MSSM can alleviate the $|V_{ub}|$ tension. The critical mechanism at work here is similar to the new flavor changing coupling in that mixing between the stop and sup squarks can be large as it is relatively unconstrained and leads to an otherwise absent CKM enhancement. The extended Isidori-Kamenik model was also shown to have the ability to alleviate the $|V_{ub}|$ tension through a flavor changing coupling between stop squarks and up quarks leading to the same CKM enhancement. We have also noted that the addition of the new flavor changing coupling associated with the $t \bar{t}$ asymmetry in conjunction with the flavor diagonal $b \bar{b}$ coupling generates a new signal which can be searched for at hadron colliders. If searches are unsuccessful then the associated bounds on the cross section stand to rule out the proposed possible connection between the $t \bar{t}$ asymmetry and the $|V_{ub}|$ tension.

\begin{center}
{\bf Acknowledgements}
\end{center}

We thank Abhishek Kumar, David Morrissey, and Sean Tulin for valuable comments and helpful discussions. Both P.W. and J.N. are supported by NSERC of Canada.

\end{document}